\documentclass[a4paper]{article}
\pdfoutput=1
\usepackage[T1]{fontenc}
\usepackage{jheppub}
\usepackage{bm}
\usepackage{graphicx}
\usepackage{subcaption}

\def\d{\delta}

\renewcommand{\Im}{\,\text{Im}\:}
\renewcommand{\Re}{\,\text{Re}\:}

\newcommand{\beq} {\begin{equation}}
\newcommand{\eeq} {\end{equation}}

\title{Non-equilibrium scalar two point functions in AdS/CFT}
\author[a]{Ville Ker\"{a}nen%
}
\author[a,b]{and Philipp Kleinert%
}
\affiliation[a]{Rudolf Peierls Centre for Theoretical Physics, University of Oxford,\\ 1 Keble Road, Oxford OX1 3NP}
\affiliation[b]{Merton College, University of Oxford,\\ Merton Street, Oxford OX1 4JD}
\emailAdd{vkeranen1@gmail.com}
\emailAdd{philipp.kleinert@physics.ox.ac.uk}

\abstract{In the first part of the paper, we discuss different versions of the AdS/CFT dictionary out of equilibrium. We show that the Skenderis--van Rees prescription and the "extrapolate" dictionary are equivalent at the level of "in-in" two point functions  of free scalar fields in arbitrary asymptotically AdS spacetimes.  In the second part of the paper, we calculate two point correlation functions in dynamical spacetimes using the "extrapolate" dictionary. These calculations are performed for conformally coupled scalar fields in examples of spacetimes undergoing gravitational collapse, the AdS$_2$-Vaidya spacetime and the AdS$_3$-Vaidya spacetime, which allow us to address the problem of thermalization following a quench in the boundary field theory. The computation of the correlators is formulated as an initial value problem in the bulk spacetime. Finally, we compare our results for AdS$_3$-Vaidya to results in the previous literature obtained using the geodesic approximation and we find qualitative agreement.}


\begin{document}

\maketitle

\newpage

\setlength{\parskip}{4mm}

\section{Introduction}

The AdS/CFT duality relates the dynamics of strongly coupled non-equilibirium quantum field theory to semiclassical gravitational dynamics \cite{Hubeny:2010ry,DeWolfe:2013cua}. By studying the duality in non-equilibrium settings, one can approach the long-standing problem of non-equilibrium quantum field theory. The gravitational description is particularly well suited for addressing questions of non-equilibrium physics since the bulk physics is classical at leading order in the $1/N$ expansion and the real time dynamics is reduced to solving differential equations, with initial data specified by the initial state in question. Using the AdS/CFT dictionary, the one point functions of boundary theory operators can be extracted using the asymptotics of the classical bulk solutions. Studying the dynamics of the one point function $\langle T_{\mu\nu}\rangle$ has lead to interesting results in the dual field theory, such as the fast thermalization of classes of out of equilibrium initial conditions and early applicability of a hydrodynamic description in strongly coupled large $N$, $\mathcal{N}=4$ super-Yang-Mills theory \cite{Chesler:2008hg,Heller:2011ju}.

The problem of thermalization is by now well studied in the condensed matter literature and it is known that the thermalization of one point functions differs in a qualitative way from thermalization of higher point correlation functions \cite{Calabrese:2006rx}. Even though expectation values of observables localized in a region of space might have reached their thermal values, the quantum correlations between different regions might still be non-thermal. Thus, in order to obtain a better understanding of thermalization in the field theories with gravitational duals, one can study how two point functions approach their thermal values. 

One question to address before this is what kind of physical observables are sensitive to the higher point correlation functions? The one point functions tell us the expectation values of physical quantities in the quantum state in question.
For example one can study the time evolution of the energy momentum tensor $\langle T_{\mu\nu}\rangle$ or an order parameter of some symmetry breaking $\langle \mathcal{O}\rangle$ (see e.g.\ \cite{Murata:2010dx,Bhaseen:2012gg,Adams:2012pj,Sonner:2014tca,Chesler:2014gya,Das:2014lda,Ewerz:2014tua}). Since we are dealing with quantum mechanical systems, the actually measured values of the observables fluctuate. These fluctuations are quantified by higher point correlation functions. For example, the temperature one measures in a subregion of a system
has fluctuations that can be quantified by the energy-momentum tensor two point function \cite{Balatsky:2014lsa}. On the other hand, (Wightman) two point functions tell us about particle creation rates, such as the photon creation rate of the quark-gluon plasma, studied in the holographic context in \cite{CaronHuot:2006te,Baier:2012ax}. As an even simpler example, the two point function of a quantum field $\phi$ quantifies the response of a detector coupled to the field $\phi$, as is familiar, for example, from the Unruh effect \cite{Unruh:1976db}. From the point of view of out of equilibrium physics in condensed matter systems, one can even directly measure real space two (and higher) point correlation functions and how they approach their thermal limits under time evolution in ultracold atomic gases \cite{Cheneau2011,Langen2013}. 

As is well known, connected two point functions are suppressed in the $1/N$ expansion compared to their disconnected parts. Thus, from the bulk point of view they are quantum mechanical. The procedure for calculating them in the bulk is the following. To leading order in $1/N$ one has a classical bulk solution for the metric and all other bulk fields. Then, one considers fluctuations around these classical backgrounds that have to be treated quantum mechanically (see e.g.\ \cite{Jackiw:1977yn} for a review of the semiclassical approximation in the context of quantum field theory). In particular, one must specify a state for these fluctuations.\footnote{In a large part of the literature, the problem of specifying a state in the bulk has been bypassed by considering retarded correlation functions, which are independent of the state in the quadratic approximation in bulk fluctuations.} In this work, we will specialize to a set of initial states that can be prepared with a Euclidean path integral over some manifold, that can be glued to the full Lorentzian spacetime.

Next, one should ask how the bulk initial state is related to the boundary theory initial state, and what is the correct dictionary between boundary and bulk quantities. Building on earlier work \cite{Maldacena:2001kr,Herzog:2002pc,Marolf:2004fy}, Skenderis and van Rees (SvR) have constructed a machinery for obtaining boundary theory correlation functions by constructing a holographic version of the Schwinger-Keldysh real time formalism \cite{Skenderis:2008dh,Skenderis:2008dg}. In their prescription, the initial state of the boundary theory is prepared by a path integral on a Euclidean manifold $M_{\partial}$, while the bulk state is prepared by the path integral over the bulk Euclidean manifolds $M$ whose boundary is $M_{\partial}$. Correlation functions are then obtained by taking functional derivatives of the on-shell action on the full glued manifold consisting of the Euclidean manifold $M$ and a Lorentzian part for the real time evolution .

Another, a priori independent dictionary between bulk and boundary correlators is provided by the "extrapolate" version of the AdS/CFT dictionary, where one obtains boundary correlation functions as boundary limits of bulk to bulk correlation functions \cite{Banks:1998dd}. In Euclidean time, the "extrapolate" dictionary is equivalent to the standard  dictionary where one takes functional derivatives of the on-shell action, as shown in \cite{Harlow:2011ke}, for
scalar fields.

In the first part of this paper, we will show that the "extrapolate" dictionary, when generalized to real time (in an obvious way), is equivalent to the SvR dictionary, for "in-in" two point correlators of a free scalar field in an arbitrary asymptotically AdS spacetime. We will demonstrate this by directly constructing a solution to the SvR equations of motion in the glued spacetime and show that this solution is related to a boundary limit of the bulk to bulk two point function. In our opinion, the equivalence of the dictionaries can be interpreted as evidence that they provide the correct dictionary for out of equilibrium physics in AdS/CFT.

In the second part of this paper, we will consider explicit examples of out of equilibrium scalar field two point functions in spacetimes undergoing gravitational collapse. We will consider states where the boundary theory is prepared in the ground state at early times, which is dual to the bulk ground state in AdS spacetime. Then, the CFT is suddenly perturbed
out of equilibrium by some operator smeared over all space. The case where the operator is a massless scalar field dual to a marginal scalar operator in the CFT was studied in \cite{Bhattacharyya:2009uu}, where it was argued that the spacetime can be well approximated by the AdS-Vaidya spacetime. The AdS-Vaidya spacetime corresponds to pressureless matter shock collapsing
along an ingoing null trajectory, and forming a black hole. The massless scalar system was further studied using simulations of numerical gravity in \cite{Wu:2012rib,Garfinkle:2011tc}, which gave further evidence for approximating the Vaidya spacetime to model the CFT process. 
As a further example, the AdS$_4$-Vaidya spacetime is known to arise as an exact dual spacetime of a CFT process where one turns on a time dependent spatially constant electric field in one of the field theory directions \cite{Horowitz:2013mia}.

We will consider the AdS vacuum state as the initial state for the scalar field in the bulk. This initial state can in principle be obtained by performing a Euclidean path integral over Euclidean AdS. However, this is not essential for us since the ground state wavefunctional of a scalar field in AdS is explicitly known. As we will review later, knowing the wavefunctional of a free field is equivalent to knowing the equal time two point function of the field. To study how the correlation functions evolve in time, we use the fact that the two point function satisfies the Klein--Gordon equation. This allows us to evolve the two point function in time by solving the Klein--Gordon equation. As the two point function depends on two bulk points $x_1$ and $x_2$, we must solve the equations of motion with respect to both of these coordinates. 

This paper is structured as follows. In Section 2, we discuss the AdS/CFT dictionary in out of equilibrium situations. In Section 3, we show how the calculation of boundary two point functions can be written as a bulk initial value problem using the "extrapolate" dictionary. In Section 4, we review the Schwinger-Keldysh formalism and the SvR prescription to calculate out of equilibrium correlation functions and show that it is equivalent to the "extrapolate" dictionary. In Section 5, we consider the example of a massless scalar in the AdS$_2$-Vaidya spacetime. In Section 6, we study a conformally coupled scalar in AdS$_3$-Vaidya spacetime. In Section 7, we compare the results of our computations in AdS$_3$-Vaidya to earlier results obtained using the geodesic approximation.

\section{On the holographic dictionary out of equilibrium}

In our practical calculations, we will use the "extrapolate" version of the AdS/CFT dictionary \cite{Banks:1998dd,Giddings:1999qu}. It directly relates correlation functions in the bulk to correlation functions of the dual
boundary operators. For example, for a scalar field of mass $m$, that is dual to a scalar operator $\mathcal{O}$ of dimension $\Delta=(d-1)/2+\sqrt{(d-1)^2/4+m^2}$, the dictionary in Euclidean time is
\beq
\langle \mathcal{O}(x_1)\ldots \mathcal{O}(x_n)\rangle_{CFT} =C_n \lim_{z\rightarrow 0}z^{-n\Delta}\langle\phi(x_1,z)\ldots \phi(x_n,z)\rangle_{bulk},\label{eq:bdhm}
\eeq
where $C_n$ is a constant related to the normalization of the CFT operators. The more conventional version of the AdS/CFT dictionary identifies the boundary generating functional of connected correlators with the Euclidean on-shell action of the bulk theory \cite{Gubser:1998bc,Witten:1998qj}. Then, correlation functions are obtained by differentiating the generating functional of connected correlators
\beq
W_E\left[\phi_0\right]= S_E^\text{on-shell}\left[\lim_{z\to 0}z^{-\Delta_-}\phi(x,z)=\phi_0(x)\right]
\eeq
according to
\beq
\langle \mathcal{O}(x_1)\ldots \mathcal{O}(x_n)\rangle_{CFT}=\tilde{C}_n\left.\frac{\delta^n}{\delta\phi_0(x_1)\ldots \delta\phi_0(x_n)} W_E\left[\phi_0\right]\right|_{\phi_0=0},\label{eq:diff}
\eeq
where again $\tilde{C}_n$ is a constant coefficient related to the normalization of the CFT operators and $\phi_0$ is the coefficient of the non-normalizable solution of the bulk equations of motion, i.e.\ $\phi_0(x)=\lim_{z\to0}z^{-\Delta_-}\phi(x,z)$, $\Delta_-=d-1-\Delta$. For Euclidean asymptotically AdS spacetimes, the two version of the dictionary have been shown to be equivalent \cite{Giddings:1999qu,Harlow:2011ke}. 

For the case of real time out of equilibrium correlation functions, where analytic continuation to a Euclidean spacetime is not available, the situation is less clear. In particular, it is not obvious how to generalize the "differentiate" dictionary (\ref{eq:diff}) to such situations to obtain all the correlation functions of interest. For retarded two point correlation functions, the out of equilibrium version of (\ref{eq:diff}) can be straightforwardly developed \cite{Balasubramanian:2012tu} building on previous work \cite{Son:2002sd,Iqbal:2008by}. Nevertheless, the generalization to other correlation functions, such as multi point Wightman correlation functions, is missing. For a class of states whose wavefunctionals can be obtained as Euclidean path integrals a generalization of the "differentiate" dictionary was constructed by Skenderis and van Rees (SvR) \cite{Skenderis:2008dh,Skenderis:2008dg} (see \cite{Maldacena:2001kr,Herzog:2002pc,Marolf:2004fy}, for earlier work in this direction). SvR constructed a holographic version of the Schwinger-Keldysh generating functional, from which all the desired correlation functions can be obtained.

On the other hand, the "extrapolate" dictionary (\ref{eq:bdhm}) has an obvious generalization to out of equilibrium correlation functions. First, one should choose the CFT state of interest. We will assume this state can be prepared with a path integral over some Euclidean manifold. We perform this Euclidean path integral in the bulk in a semiclassical approximation to prepare the state of the bulk quantum fields. Then, one should choose the CFT correlation function of interest and make the replacement
\beq
\mathcal{O}(x)= \lim_{z\rightarrow 0}z^{-\Delta}\phi(x,z),\label{eq:operatordict}
\eeq
in the correlation function. This gives the bulk correlation function one should compute. For example, if one wants to calculate the boundary time ordered two point function $\langle T\mathcal{O}(x_1)\mathcal{O}(x_2)\rangle$, then the substitution (\ref{eq:operatordict}) tells us we should compute the bulk correlation function
\beq
\langle T\mathcal{O}(x_1)\mathcal{O}(x_2)\rangle=C_2\lim_{z_1\rightarrow 0}\lim_{z_2\rightarrow 0}z_1^{-\Delta}z_2^{-\Delta}\langle T\phi(x_1,z_1)\phi(x_2,z_2)\rangle.\label{eq:2pointextrapolate}
\eeq
This way, the calculation of the boundary two point function becomes a standard problem of quantum field theory in curved spacetime.

This procedure can be shown to give correct results for correlation functions in the vacuum and in thermal equilibrium, as is shown for example in Appendix A of \cite{Papadodimas:2012aq} for the BTZ black hole. Issues such as choosing ingoing versus outgoing boundary conditions and the problem of boundary terms from the horizon \cite{Son:2002sd} simply do not exist in this formalism. Thus, we view the "extrapolate" dictionary as a natural starting point for out of equilibrium AdS/CFT.

\section{Two point functions as initial value problems}\label{sec:initialvalue}

Our method for calculating the bulk two point function is to view the time evolution of the correlation function as an initial value problem. In this work, we will only consider free scalar fields in the bulk. When the one point function of the scalar vanishes, this is sufficient for the purpose of calculating two point functions to leading order in the $1/N$ expansion.\footnote{If the one point function $\langle \phi\rangle$ is non-vanishing, one should include bulk Feynman/Witten diagrams, where $\phi$ mixes with the graviton and other bulk fields, which gives contributions to the two point function that are not suppressed by powers of $1/N$ from the free result. Thus, we will in the following assume that the one point function $\langle \phi\rangle$ vanishes. In our explicit examples, this follows from conformal symmetry of the initial state, and from the $\phi\rightarrow -\phi$ symmetry of the action (\ref{eq:scalaraction}).} To fix our conventions, we use the following action for a free scalar field
\beq
S=-\frac{1}{2}\int d^dx\sqrt{-g}\Big(\partial_{\mu}\phi\partial^{\mu}\phi+m^2\phi^2\Big).\label{eq:scalaraction}
\eeq
In the Heisenberg picture, the bulk scalar quantum field satisfies its Heisenberg equation of motion
\beq
(\Box-m^2)\phi(x)=0.
\eeq
Moreover, the state $|\psi\rangle$ of the system does not vary in time and can be fixed by an initial condition. A straightforward exercise shows that the bulk Feynman two point correlator
\beq
G_F(x_2,x_1)=\langle\psi| T \phi(x_2)\phi(x_1)|\psi\rangle
\eeq
satisfies
\beq
(\Box_1-m^2)G_F(x_2,x_1)=i\frac{\delta(x_2-x_1)}{\sqrt{-g}}=(\Box_2-m^2)G_F(x_2,x_1).\label{eq:KGeqs}
\eeq
Thus, if we know $G_F(x_2,x_1)$ and its first time derivative on some initial time slice, we can use the equations of motion
to evolve it forwards in time. For free bulk fields, this implies that the specification of the initial state is identical
to specifying the two point correlation function and its first time derivatives with respect to $t_1$ and $t_2$ as initial data.  

\begin{figure}
\centering
\includegraphics[scale=.8]{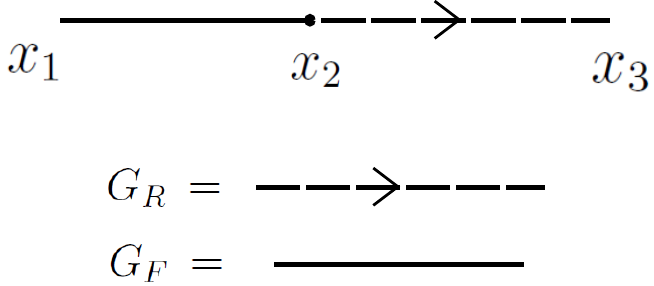}
\caption{\label{fig:joining1} The main joining formula (\ref{eq:joiningform1}) visualized in a diagrammatic form.}
\end{figure}

How do we solve the equations of motion (\ref{eq:KGeqs}) given some initial data? A standard solution to this problem is to use the method of Green's functions \cite{Ebrahim:2010ra,CaronHuot:2011dr,Balasubramanian:2012tu}
\beq
G_F(x_3,x_1)=i \int_{t_2=\textrm{const}} d\bm{x}_2\,G_F(x_2,x_1) \overleftrightarrow{D}^{t_2}G_R(x_3,x_2),\label{eq:joiningform1}
\eeq
where $D^{t}=\sqrt{-g}g^{t\mu}\partial_{\mu}$ and the integral is over all space on a constant time slice. In the following, we will use bold letters for spatial vectors. Thus, the integral measure $d\bm{x}_2=d^{d-1}x_2$ denotes an integral over a constant time slice. Here, we have chosen the times in the order $t_1<t_2<t_3$. This formula means that if we know the two point function $G_F(x_2,x_1)$ we can obtain it at some later time $t_3$ by using (\ref{eq:joiningform1}). The formula is visualized in Figure \ref{fig:joining1} in a diagrammatic form.

A derivation of (\ref{eq:joiningform1}) goes as follows
\beq
G_F(x_3,x_1)=-i\int d^dx\sqrt{-g}G_F(x,x_1)(\Box-m^2)G_R(x_3,x),\label{eq:jderivation2}
\eeq
where we used
\beq
(\Box-m^2)G_R(x_3,x)=i\frac{\delta(x_3-x)}{\sqrt{-g}}.\label{eq:retardedeom}
\eeq
We will choose the integration region in (\ref{eq:jderivation2}) to run over all space from time $t=t_2$ to some final time 
$t_f>t_3$, where $t_1<t_2<t_3$. Then, using the identity
\beq
\sqrt{-g}G_F\Box G_R=\partial_{\mu}(G_F\overleftrightarrow{D}^{\mu}G_R)+\sqrt{-g}G_R\Box G_F
\eeq
and (\ref{eq:KGeqs}), we obtain
\beq
G_F(x_3,x_1)=-i\int d\bm{x}\int_{t_2}^{t_f}dt\, \partial_{t}\Big(G_F(x,x_1) \overleftrightarrow{D}^{t}G_R(x_3,x)\Big),
\eeq
which gives (\ref{eq:joiningform1}) after noting that $G_R(x_3,x_f)\propto \theta(t_3-t_f)=0$ while the boundary term
from the AdS boundary vanishes as the two point functions approach zero with the rate $z^{\Delta}$.

A convenient fact is that the retarded correlator can be obtained from the Feynman correlator using the identity
\beq
G_R(x_2,x_1)=\theta(t_2-t_1)\langle [\phi(x_2),\phi(x_1)]\rangle=2i\theta(t_2-t_1)\,\text{Im}\,G_F(x_2,x_1).
\eeq
This identity simply follows from the operator definition of $G_F$. Some relations like this among real time correlation functions are summarized in Appendix \ref{sec:correlatoridentities}. Another convenient fact is that for a free field $G_R$ does not depend on the state of the system. This can be understood as follows. Since the operators $\phi(x)$ commute at equal time it follows that $G_R=0$ on an equal time slice. The first time derivatives of $G_R$ are proportional to equal time canonical commutators, e.g.\ $\partial_{t_2}G_R\propto \left.\langle[\phi(x_1),\Pi(x_2)]\rangle\right|_{t_1=t_2}=i\delta(\bm{x}_1-\bm{x}_2)$. Thus, also the first time derivatives of $G_R$ are independent of the state on an equal time slice. This means that the initial data for $G_R$ is independent of the state at an initial time slice and since $G_R$ satisfies (\ref{eq:retardedeom}), a second order differential equation, it is independent of the state at all times.\footnote{The correlator $G_R$ still depends on the metric of the bulk spacetime. The above simply means that $G_R$ does not depend on the bulk quantum state of the scalar field while, since it depends on the spacetime in question, it depends on the boundary CFT state.}

\begin{figure}
\centering
\includegraphics[scale=.6]{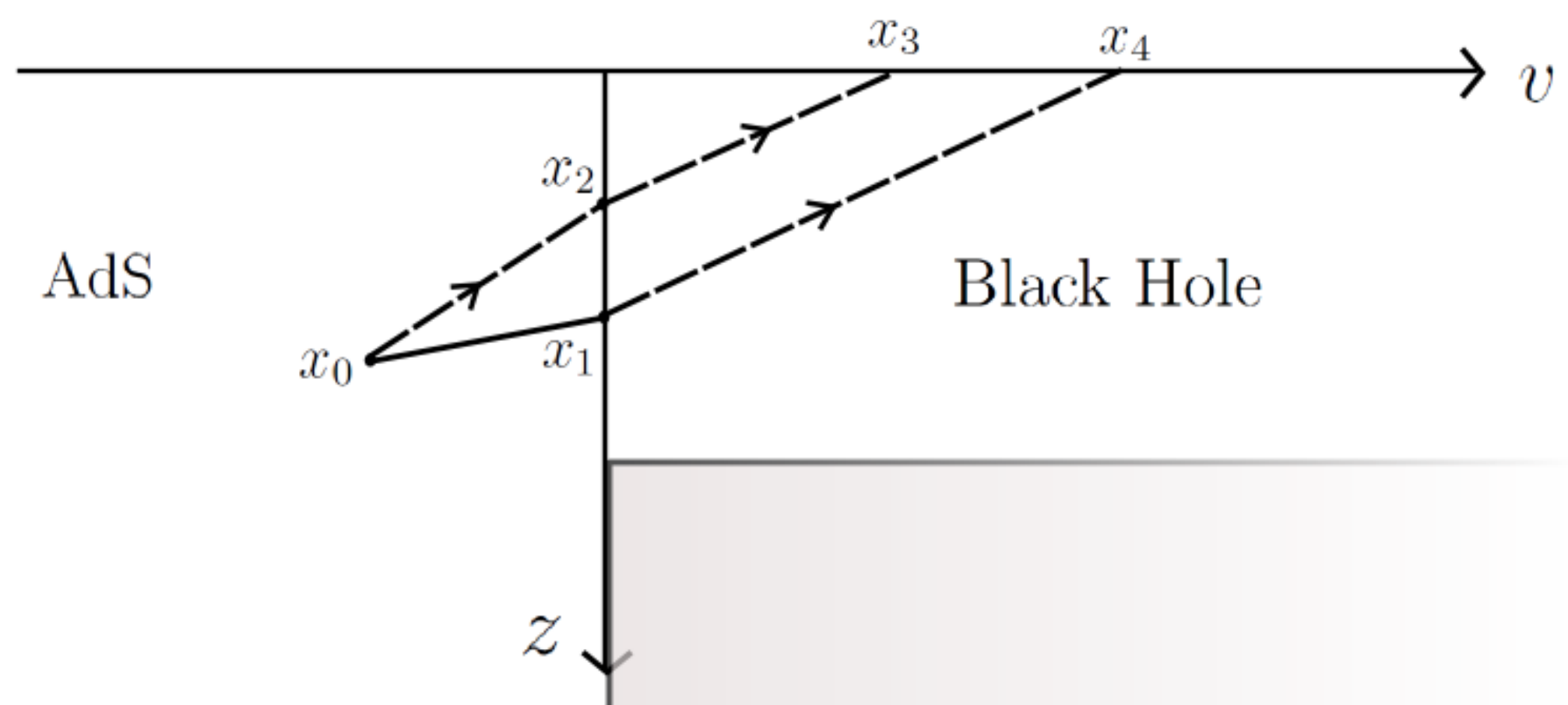}
\caption{\label{fig:joining2} The joining formula (\ref{eq:joining3propa}) we use in the AdS-Vaidya spacetime visualized in a diagrammatic form.
The grey shaded area is the interior of the apparent horizon.}
\end{figure}

The same procedure can be used to propagate the point $x_1$ in $G_F(x_3,x_1)$ forwards in time. For technical reasons that we will explain in a moment, we have found it necessary to rather use the joining formula (\ref{eq:joiningform1}) three times in the explicit calculations
\beq
G_F(x_4,x_3)=-i \int d\bm{x}_2 d\bm{x}_1 d\bm{x}_0\, \Big(G_F(x_1,x_0)\overleftrightarrow{D}^{t_1}G_R(x_4,x_1)\Big)\overleftrightarrow{D}^{t_0} \Big(G_R(x_2,x_0)\overleftrightarrow{D}^{t_2}G_R(x_3,x_2)\Big).\label{eq:joining3propa}
\eeq
This is visualized in Figure \ref{fig:joining2} in a diagrammatic form for the case where the bulk spacetime is the AdS-Vaidya spacetime. This is the basis of our numerical calculations. Also we will choose to use an ingoing null coordinate $v$ as the time coordinate, since the Vaidya spacetime is particularly simple in terms of this coordinate. It is convenient to choose the integrals $d\bm{x}_1$ and $d\bm{x}_2$ to be located at the value of $v$, when the collapse in Vaidya happens, while the integral $d\bm{x}_0$ is on some (arbitrary) earlier value of lightcone time $v$. Now, we can see why using only two joining integrals fails in the case of Vaidya spacetime. We could have instead performed two joining integrals $d\bm{x}_1$ and $d\bm{x}_2$ on the lightcone time slice coincident with the shock wave, where the initial data would then be given by the Feynman two point function $G_F(x_2,x_1)$ at the time of the infalling shock wave. The problem is that the Feynman two point function has divergences  when the points $x_1$ and $x_2$ are lightlike separated. In the case when $v_1\not=v_2$, we only encounter divergences with power $-1/2$ and one divergence with power $-1$ at a single point where two lightcone divergences merge. In contrast, for $v_1\not=v_2$ there is a line of divergences with power $-1$. Thus, we use a third joining integral at some earlier time $v_0$ to circumvent these lightcone divergences with power $-1$ in the initial data.

\section{Equivalence of two out of equilibrium dictionaries}\label{sec:dictproof}

\subsection{Review of the Schwinger-Keldysh formalism and the Skenderis--van Rees prescription}

\begin{figure}
\centering
\includegraphics[scale=.8]{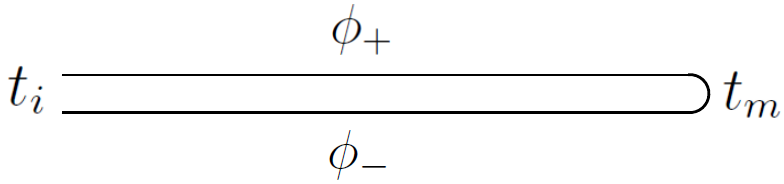}
\caption{\label{fig:ctime} The complex time contour for "in-in" correlation functions.}
\end{figure}

Since we are interested in calculating expectation values of operators in a given state $|\psi\rangle$, a standard Feynman path integral is not sufficient as it computes transition amplitudes from initial to final states. This is not suitable for the non-equilibrium setting as we are interested in studying
the time evolution of observables without presupposing a final state. To find the correct generalization to non-equilibrium situations, we study the following
two point function as an example
\beq
	\langle\psi|\mathcal{O}(t_2,\bm{x}_2)\mathcal{O}(t_1,\bm{x}_1)|\psi\rangle=\langle\psi|U^{\dagger}(t_2,t_i)\mathcal{O}(t_i,\bm{x}_2)U(t_2,t_1)\mathcal{O}(t_i,\bm{x}_1)U(t_1,t_i)|\psi\rangle,\label{eq:SK1}
\eeq
where $U(t_2,t_1)$ is the time evolution operator from $t_1$ to $t_2$. For concreteness, let us suppose that $t_2>t_1$. The operator $\mathcal{O}$ is some local operator built out of the fields in the theory $\varphi$. A standard time slicing argument (see e.g.\ \cite{Kleinert:2009}) shows that the time evolution operators may be written in a path integral form as
\beq
	U(t_2,t_1)=\int[d\varphi]\,|\varphi(t_2)\rangle e^{iS[\varphi]}\langle\varphi(t_1)|,\label{eq:timegen}
\eeq
where $|\varphi\rangle$ is a field operator eigenstate. Replacing the time evolution operators in (\ref{eq:SK1}) with (\ref{eq:timegen}), we end up with the path integral
\begin{align}
	\langle\psi|\mathcal{O}(t_2,\bm{x}_2)&\mathcal{O}(t_1,\bm{x}_1)|\psi\rangle\notag\\
		=&\int [d\varphi_+ d\varphi_-]\,\Psi^*[\varphi_-(t_i)]\Psi[\varphi_+(t_i)]e^{iS[\varphi_+]-iS[\varphi_-]}\mathcal{O}(x_1)\mathcal{O}(x_2)\delta(\varphi_+(t_m)-\varphi_-(t_m)),
\end{align}
where we introduced the wavefunctional $\Psi[\varphi]=\langle \varphi|\psi\rangle$. The $\varphi_+$ path integral corresponds to time evolution forwards in time until the time of the latest operator,\footnote{Of course, one can also time evolve to later times than the latest operator, but then the forwards time evolution is cancelled by the backwards time evolution by unitarity.} while the path integral over $\varphi_-$ corresponds to time evolution backwards in time towards the initial time. Thus, one can imagine that the path integral is over fields living on a closed time contour, which runs from the initial time $t_i$ to some time $t_m$ later than the last operator insertion and then back to $t_i$. This time contour is visualized in Figure \ref{fig:ctime}.

Correlation functions can also be computed by functional differentiation of the following Schwinger-Keldysh generating functional
\beq
	Z[J_+,J_-]=\int [d\varphi_+ d\varphi_-]\,\Psi^*[\varphi_-(t_i)]\Psi[\varphi_+(t_i)]e^{iS[\varphi_+] - iS[\varphi_-]+i\int(\mathcal{O}_+J_+-\mathcal{O}_- J_-)}\,\delta(\varphi_+(t_m)-\varphi_-(t_m)),
\eeq
This generating functional calculates correlation functions that are ordered along the time contour in Figure \ref{fig:ctime}.
Continuing earlier work \cite{Maldacena:2001kr,Herzog:2002pc,Marolf:2004fy}, SvR suggested that the bulk dual of the field theory Schwinger-Keldysh generating functional is given by the bulk path integral \cite{Skenderis:2008dh,Skenderis:2008dg}
\beq
	Z[J_+,J_-]=\int [d\Phi_+ d\Phi_-]e^{iS[\Phi_+]-iS[\Phi_-]}\Psi^*[\Phi_-(t_i)]\Psi[\Phi_+(t_i)]\delta(\Phi_+(t_m)-\Phi_-(t_m)),\label{eq:svrpathintegral}
\eeq
where the boundary sources are, as usual, identified as the asymptotic values of bulk fields
\beq
	\Phi_{\pm}=z^{\Delta_-}J_{\pm}(x)+\ldots\;,
\eeq
where $\Phi$ collectively denotes all the bulk fields. 
Using the semiclassical approximation in the bulk, the path integral (\ref{eq:svrpathintegral}) becomes
\beq
	Z[J_+,J_-]\approx e^{iS[\Phi_+]-iS[\Phi_-]}\Psi^*[\Phi_-(t_i)]\Psi[\Phi_+(t_i)],\label{eq:svrpathintegral2}
\eeq
where the bulk fields $\Phi_{\pm}$ satisfy the saddle point condition
\beq
	\frac{\delta}{\delta\Phi_{\pm}}\Bigg[iS[\Phi_+]-iS[\Phi_-]+\log\Psi^*[\Phi_-(t_i)]+\log\Psi[\Phi_+(t_i)]\Bigg]=0\label{eq:SvRalleoms}
\eeq
Furthermore, SvR considered the case where the bulk initial state $\Psi$ is obtained from a path integral over Euclidean manifolds, that can be glued in to the Lorentzian part of the bulk spacetime. The semiclassical approximation reduces this path integral to a single Euclidean manifold. Then, (\ref{eq:SvRalleoms}) gives rise to the equations of motion for all the bulk fields and conditions for how the fields are glued to the Euclidean manifold, at time $t_i$. In the following, we will assume that the metric and possible other fields are consistently glued. The details of the gluing procedure can be found in \cite{Skenderis:2008dg}.

One of the bulk fields $\Phi$ is the scalar field $\phi$, with the action (\ref{eq:scalaraction}).
We are interested in calculating correlation functions of the CFT operator dual to $\phi$.
Thus, our task is to solve (\ref{eq:SvRalleoms}) for the scalar field.
The wavefunctional of the scalar field in the semiclassical approximation is given by the on-shell Euclidean action (see e.g. \cite{Jackiw:1987ar,Weinberg:1995mt,Witten:2001kn} for discussions on wavefunctionals in quantum field theory)
\beq
	\Psi[\phi_i]\approx \frac{1}{\sqrt{\mathcal{N}}}e^{-S_E[\phi_E]}\Big|_{\phi_E(t_i)=\phi_i}=\frac{1}{\sqrt{\mathcal{N}}}e^{-\frac{1}{4}\int d\bm{x}_1\,d\bm{x}_2\,\phi_i(\bm{x}_1)K(\bm{x}_1,\bm{x}_2)\phi_i(\bm{x}_2)},
	\label{eq:wavefunctional}
\eeq
where we used the fact that the on-shell action can be written as a boundary term at the gluing surface and that it is a quadratic functional of $\phi_i$ after using the Euclidean equations of motion.\footnote{The reader unfamiliar with this fact should consult Problem 3, in Section F of \cite{Harlow:2014yka}.} The last equality in (\ref{eq:wavefunctional}) defines $K$ in terms of the Euclidean manifold and the action functional. This wavefunctional is defined on an initial Cauchy slice, i.e.\ on a spatial slice $t=t_i$, which we assume to exist. Above, $\mathcal{N}$ is a normalization factor chosen so that the norm of the wavefunction is one, i.e.
\beq
	\mathcal{N}=\int [d\phi_i]e^{-\frac{1}{2}\int d\bm{x}_1\,d\bm{x}_2\,\phi_i(\bm{x}_1)K(\bm{x}_1,\bm{x}_2)\phi_i(\bm{x}_2)},
\eeq
where the integral over field configurations integrates over spatial configurations at fixed time $t=t_i$. The kernel $K(\bm{x}_1,\bm{x}_2)$ is related to the initial equal time two point function of the bulk field $\phi$ through
\beq
	G(\bm{x}_1,\bm{x}_2)=\left.\langle\psi| \phi(x_1)\phi(x_2)|\psi\rangle \right|_{t_1=t_2=t_i}=\frac{1}{\mathcal{N}}\int [d\phi_i]e^{-\frac{1}{2}\int \phi_i K\phi_i}\phi_i(\bm{x}_1)\phi_i(\bm{x}_2),
\eeq
where
\beq
	\int \phi_i K\phi_i=\int d\bm{x}_1\,d\bm{x}_2\,\phi_i(\bm{x}_1)K(\bm{x}_1,\bm{x}_2)\phi_i(\bm{x}_2).
\eeq
The path integral is Gaussian and can be easily performed to give
\beq
	G(\bm{x}_1,\bm{x}_2)=K^{-1}(\bm{x}_1,\bm{x}_2).
\eeq
Thus, $K$ is simply the inverse of the initial equal time two point function and is implicitly determined in terms of $G$ by
\beq
	\int d\bm{x}\, G(\bm{x}_1,\bm{x})K(\bm{x},\bm{x}_2)=\delta(\bm{x}_1-\bm{x}_2).\label{eq:inverseK}
\eeq
In the following, we will not need the explicit form of the kernel $K(\bm{x}_1,\bm{x}_2)$, except that we will assume that it is a real function which satisfies (\ref{eq:inverseK}).

For the scalar field, the saddle point condition (\ref{eq:SvRalleoms}) becomes
\beq\label{eq:variation}
\frac{\delta}{\delta \phi_{\pm}}\Big(iS[\phi_+]-iS[\phi_-]-\frac{1}{4}\int \phi_+ K\phi_+-\frac{1}{4}\int \phi_- K\phi_-\Big)=0,
\eeq
where 
\begin{align}
	\int\phi_+ K\phi_+&=\int d\bm{x}_1\int d\bm{x}_2 \,\phi_+(\bm{x}_1,t_i) K(\bm{x}_1,\bm{x}_2)\phi_+(\bm{x}_2,t_i),\\
	\int \phi_- K\phi_-&=\int d\bm{x}_1\int d\bm{x}_2\, \phi_-(\bm{x}_1,t_i) K(\bm{x}_1,\bm{x}_2)\phi_-(\bm{x}_2,t_i).
\end{align}
The variation of (\ref{eq:variation}) with respect to $\phi_{\pm}$ away from the endpoints of the contour leads to the bulk equation of motion
\beq
	(\Box-m^2)\phi_{\pm}=0.\label{eq:svreom}
\eeq
The variation of $\phi_+$ at the initial time $t_i$ gives an initial condition
\beq
	D^t\phi_+(\bm{x},t)|_{t=t_i}=-\frac{i}{2}\int d\bm{x}_1 \phi_+(\bm{x}_1,t_i) K(\bm{x}_1,\bm{x}),\label{eq:ini1}
\eeq
where $D^t=\sqrt{-g}g^{t\mu}\partial_{\mu}$.
Similarly, the variation of $\phi_-$ at the initial time gives an initial condition for $\phi_-$,
\beq
	D^t\phi_-(\bm{x},t)|_{t=t_i}=\frac{i}{2}\int d\bm{x}_1 \phi_-(\bm{x}_1,t_i) K(\bm{x}_1,\bm{x}),\label{eq:ini2}
\eeq
which differs from the boundary condition for $\phi_+$ by a single minus sign. Variation at the "turning point" of the contour, $t=t_m$, gives
\beq
	D^t\phi_+(\bm{x},t)|_{t=t_m} = D^t\phi_-(\bm{x},t)|_{t=t_m}.
	\label{eq:mid1}
\eeq
The delta functional in the original path integral gives one more condition at $t=t_m$,
which is the continuity of the fields
\beq
	\phi_+(\bm{x},t)|_{t=t_m} =\phi_-(\bm{x},t)|_{t=t_m}.\label{eq:mid2}
\eeq
Thus, in the SvR prescription, we have to solve the equations of motion (\ref{eq:svreom}) for $\phi_+$ and $\phi_-$ with the boundary conditions (\ref{eq:ini1}), (\ref{eq:ini2}), (\ref{eq:mid1}) and (\ref{eq:mid2}).

\subsection{Proof of equivalence of the two dictionaries}

Instead of finding a general solution to the set of equations, that is obtained in the SvR prescription, we start from an ansatz motivated by the "extrapolate" dictionary and show that it solves all the equations (\ref{eq:svreom}), (\ref{eq:ini1}), (\ref{eq:ini2}), (\ref{eq:mid1}) and (\ref{eq:mid2})

As in Euclidean time AdS/CFT, the most general solution to the bulk equations of motion, satisfying the boundary conditions
\beq
	\phi_{\pm}(x_B,z)=z^{\Delta_-} J_{\pm}(x_B)+\ldots\;,
\eeq
can be written in terms of bulk to boundary propagators
\begin{align}
	\phi_+(x)&=\int d\bm{x}_B\, K_{++}(x,x_B) J_+(x_B)+\int d\bm{x}_B\, K_{+-}(x,x_B) J_-(x_B),\label{eq:bulk_to_boundary_plus}\\
	\phi_-(x)&=\int d\bm{x}_B\, K_{-+}(x,x_B) J_+(x_B)+\int d\bm{x}_B\, K_{--}(x,x_B) J_-(x_B),\label{eq:bulk_to_boundary_minus}
\end{align}
where $x_B$ is a point at the boundary. In the SvR prescription, the boundary theory
correlation functions are determined by taking functional derivatives of the generating functional $\log Z$, i.e. 
of the on-shell action. This way, the boundary two point functions of interest are given by
\begin{align}
	&\langle T \mathcal{O}(x_1)\mathcal{O}(x_2)\rangle =-2i\nu K_{++}^{(1)}(x_1,x_2),\label{eq:svrdict1}\\
	&\langle \mathcal{O}(x_1)\mathcal{O}(x_2)\rangle =-2 i\nu K_{-+}^{(1)}(x_1,x_2),\label{eq:svrdict2}
\end{align}
where 
\beq\label{eq:bbcond}
K_{\alpha\beta}(x_1,z_1;x_2)=z_1^{\Delta_-}\delta_{\alpha\beta}\delta^{d-1}(x_1-x_2)+z_1^{\Delta}K_{\alpha\beta}^{(1)}+\ldots\;.
\eeq
The other correlation functions can be obtained from these by taking complex conjugates. 
We will first consider the case where $J_-=0$ and $J_+\neq 0$, in
which case we need to construct the bulk to boundary propagators $K_{++}$ and $K_{-+}$. The opposite case
 $J_-\neq 0$ and $J_+= 0$ can be obtained from the former one, by performing a time reversal transformation on
the complex time contour. Then, due to the linearity of the problem, the general case is a superposition of 
these two.

\paragraph{Claim:} $K_{++}$ and $K_{-+}$ are given by boundary limits of the following bulk correlators
\begin{align}
	K_{++}(x,x_B)&=2i\nu \lim_{z_B\rightarrow 0}z_B^{-\Delta}G_F(x;x_B,z_B),\label{eq:claim1}\\
	K_{-+}(x,x_B)&=2i\nu\lim_{z_B\rightarrow 0}z_B^{-\Delta}G_+(x;x_B,z_B),\label{eq:claim2}
\end{align}
where we denote $\nu=(\Delta-\Delta_-)/2=\sqrt{(d-1)^2/4+m^2}$ and
\beq
	G_F(x_1,x_2)=\langle  \psi|T \phi(x_1)\phi(x_2)|\psi\rangle,\quad G_+(x_1,x_2)=\langle \psi|\phi(x_1)\phi(x_2)|\psi\rangle,
\eeq
which are the bulk Feynman and Wightman functions, computed in the state $|\psi\rangle$. Thus, they satisfy the
equations of motion
\beq
	(\Box-m^2)G_F(x_1,x_2)=i\frac{\delta^d(x_1-x_2)}{\sqrt{-g}},\quad (\Box-m^2)G_+(x_1,x_2)=0,\label{eq:Gfeom1}
\eeq
with respect to both of their arguments. Also they satisfy the initial conditions
\beq
	\left.G_F(x_1,x_2)\right|_{t_1=t_2=t_i}=G(x_1,x_2),\quad \left.G_+(x_1,x_2)\right|_{t_1=t_2=t_i}=G(x_1,x_2).\label{eq:ini3}
\eeq

\paragraph{Proof:} We want to show that $K_{++}$ and $K_{-+}$, constructed using the "extrapolate" dictionary, (\ref{eq:claim1}) and (\ref{eq:claim2}), are equivalent to the bulk to boundary propagators obtained using the SvR prescription. The equations determining the bulk fields in the SvR prescription, namely the equations of motion (\ref{eq:svreom}) for $\phi_+$ and $\phi_-$ with the boundary conditions (\ref{eq:ini1}), (\ref{eq:ini2}), (\ref{eq:mid1}) and (\ref{eq:mid2}), translate into conditions for the bulk to boundary propagators using (\ref{eq:bulk_to_boundary_plus}) and (\ref{eq:bulk_to_boundary_minus}). We show that (\ref{eq:claim1}) and (\ref{eq:claim2}) satisfy these conditions and the correct boundary conditions (\ref{eq:bbcond}).

First, consider the initial condition (\ref{eq:ini1}). We have to show that
\beq
	\left.D^t G_F(x,x_B)\right|_{t=t_i}=-\frac{i}{2}\int d\bm{x}_1\left.G_F(x_1,x_B)  K(\bm{x}_1,\bm{x})\right|_{t_1=t_i}.
\label{eq:ini1_1}
\eeq
Since $G_F(x,x_B)$ satisfies the Klein--Gordon equation with a delta function source, we can use a "joining formula"
\beq
	G_F(x_2,x_B)=\theta(t_2-t_1)G_R(x_B,x_2)+i\int d\bm{x}_1\, G_F(x_2,x_1) \overleftrightarrow{D}^{t_1}G_R(x_B,x_1),
\eeq
where we choose $t_1=t_i+\epsilon$ and the integral is performed on a constant time slice. We need to separate $t_i$ and $t_1$ a bit to make sense of the distributions that appear in the following manipulations. The final result is still unambiguous. To check (\ref{eq:ini1_1}), we need to take a time derivative
\beq
	D^{t_2}G_F(x_2,x_B)=D^{t_2}(\theta(t_2-t_1)G_R(x_B,x_2))+i\int d\bm{x}_1\, D^{t_2} G_F(x_2,x_1) \overleftrightarrow{D}^{t_1}G_R(x_B,x_1).
\eeq
To proceed, we need the derivatives $D^{t_2}G_F(x_2,x_1)|_{t_2=t_1+\epsilon}$ and $D^{t_1}D^{t_2}G_F(x_2,x_1)|_{t_2=t_1+\epsilon}$.
These can be calculated by using path integrals and the initial wavefunctional since time derivatives of the field are given by  the conjugate momentum $\Pi=-D^t\phi$ and the conjugate momentum operator in the path integral becomes a functional derivative with respect to the conjugate field. These path integrals are performed in Appendix \ref{sec:pathintegrals} and the results are
\begin{align}
	D^{t_2}G_F(x_2,x_1)|_{t_2=t_1+\epsilon} &= \frac{i}{2}\textrm{sign}(\epsilon)\delta (\bm{x}_1-\bm{x}_2)+O(\epsilon),\label{eq:inideriv}\\
	D^{t_1}D^{t_2}G_F(x_2,x_1)|_{t_2=t_1+\epsilon} &= \frac{1}{4}K(x_1,x_2)-iD^{t_2}\theta(t_2-t_1)\delta(\bm{x}_1-\bm{x}_2)|_{t_2=t_1+\epsilon}+O(\epsilon).
\end{align}
The above results can be also obtained directly from the Klein--Gordon equation, but we find the path integral derivation
more convenient. Using these we obtain
\beq\label{eq:rhs}
	\left.D^{t_2}G_F(x_2,x_B)\right|_{t_2=t_i}=\frac{1}{2}D^{t_2}G_R(x_B,x_2)-\frac{i}{4}\int d\bm{x}_1\, K(\bm{x}_2,\bm{x}_1)G_R(x_B,x_1),
\eeq
where all the terms involving $\epsilon$ have disappeared, showing that the result is independent of the regulator.
This is the left hand side of (\ref{eq:ini1_1}). The right hand side of (\ref{eq:ini1_1}) is then given by
\beq\label{eq:lhs}
	-\frac{i}{2}\int d\bm{x}_1 K(\bm{x}_1,\bm{x}_2)\Big(\theta(t_1-t_3)G_R(x_B,x_1)+i\int d\bm{x}_3\,G_F(x_1,x_3)\overleftrightarrow{D}^{t_3}G_R(x_B,x_3)\Big),
\eeq
where we again used a joining integral to express $G_F$ as an integral over an initial time slice at $t_3=t_1+\delta$. Using (\ref{eq:inideriv}) and $\int d\bm{x}_1 \left.K(\bm{x}_1,\bm{x}_2)G_F(x_1,x_3)\right|_{t_1=t_3=t_i}=\delta(\bm{x}_2-\bm{x}_3)$, which follows from the initial condition for $G_F$ (\ref{eq:ini3}) and from (\ref{eq:inverseK}), it is easy to see that (\ref{eq:lhs}) is identical to (\ref{eq:rhs}). This proves that (\ref{eq:claim1}) satisfies the correct initial condition (\ref{eq:ini1}).

The same calculation can be repeated for $G_+(x,x_B)$. According to (\ref{eq:ini2}), we have to show that
\beq
	\left.D^t G_+(x,x_B)\right|_{t=t_i}=-\frac{i}{2}\int d\bm{x}_1\left.G_+(x_1,x_B)  K(\bm{x}_1,\bm{x})\right|_{t_1=t_i}.
\label{eq:ini2_1}
\eeq
According to the results of Appendix \ref{sec:pathintegrals}, the early time derivatives are given by
\begin{align}
	\left.D^{t_2}G_+(x_2,x_1)\right|_{t_1=t_2=t_i}&=\frac{i}{2}\delta(\bm{x}_1-\bm{x}_2)+O(\epsilon)=-\left.D^{t_1}G_+(x_2,x_1)\right|_{t_1=t_2=t_i},\label{eq:inideriv2}\\
	\left.D^{t_1}D^{t_2}G_+(x_2,x_1)\right|_{t_1=t_2=t_i}&=\frac{1}{4}K(\bm{x}_1,\bm{x}_2)+O(\epsilon),\label{eq:inideriv3}
\end{align}
and the joining formula is given by
\beq
	G_+(x_,x_B)=i\int d\bm{x}_1\,G_+(x,x_1)\overleftrightarrow{D}^{t_1}G_R(x_B,x_1),
\eeq
where the integral is performed over the constant time slice at $t_1=t_i$. Plugging the joining formula into (\ref{eq:ini2}) and using (\ref{eq:inideriv2}) and (\ref{eq:inideriv3}), it is easy to see that (\ref{eq:ini2_1}) is satisfied. 

Thus, we have shown that (\ref{eq:claim1}) and (\ref{eq:claim2}) both satisfy the correct initial conditions. It is clear that (\ref{eq:claim2}) satisfies the right equation of motion, while (\ref{eq:claim1}) satisfies the right equation of motion everywhere except at the point near the boundary $x=x_B$, where there is a delta function. We will study this point closer in a moment.

Both of the bulk to bulk correlators in (\ref{eq:claim1}) and (\ref{eq:claim2}) satisfy normalizable boundary conditions at the AdS boundary, except possibly at $x=x_B$, where the equation of motion for the Feynman propagator has a singularity. Thus, we are lead to conclude that  (\ref{eq:claim1}) and (\ref{eq:claim2}) satisfy the correct boundary conditions (\ref{eq:bbcond}) at the AdS boundary except possibly at $x=x_B$, where the bulk to boundary correlator has a delta function non-normalizable mode. Above, we assumed the well-known fact that, given a solution with a vanishing non-normalizable part, time evolution according to the unsourced equation of motion will not generate one. Our next task is to show that (\ref{eq:claim1}) also has a delta function non-normalizable mode  at $x=x_B$. This follows directly from the delta function on the right hand side of the equation of motion (\ref{eq:Gfeom1}) for $G_F$ (see e.g. \cite{CaronHuot:2011dr} for a similar proof on a bulk string worldsheet). To see this, we consider the equation of motion in the region of small $z$, where the geometry is well approximated by AdS
\beq
	(\Box-m^2)G_F(x,x_B)=\Big(z^{d}\partial_z(z^{2-d}\partial_z)+z^2(\partial_i^2-\partial_t^2)-m^2\Big)G_F(x,x_B)=iz^{d}\delta^d(x-x_B).
\eeq
At small $z$, the derivative terms in the boundary directions are subleading and can be neglected. Rewriting the $z$ derivatives in a convenient way and multiplying the equation with $z^{\Delta-d}$ in both sides, we obtain
\beq
	\partial_z\Big[z^{2\Delta+2-d}\partial_z\Big(z^{-\Delta}G_F(x,x_B)\Big)\Big]= i\,z^{\Delta}\delta^d(x-x_B).
\eeq
Integrating both sides over $z$, from zero to $z>z_B$, and using the fact that for $z< z_B$, the Feynman
correlator satisfies the normalizable boundary conditions, leads to
\beq
	z^{2\Delta+2-d}\partial_z\Big(z^{-\Delta}G_F(x,x_B)\Big)= iz_B^{\Delta}\delta(x-x_B).\label{eq:deltafunc1}
\eeq
While $G_F(x,x_B)$ decays as $z^{\Delta}$ for $z\ll z_0$, it can have a part which decays as $z^{\Delta_-}$,
for $z>z_B$. Denoting this part as
\beq
	G_F(x,x_B)=z^{\Delta_-}G_F^{\partial}(x,x_B)+\ldots \,,\quad \textrm{for}\quad z>z_B,
\eeq
and plugging into (\ref{eq:deltafunc1}), gives
\beq
	-2\nu G_F^{\partial}(x,x_B)=iz_B^{\Delta}\delta^{d-1}(x-x_B).
\eeq
Thus, as long as we first take $z_B\rightarrow 0$, before $z\rightarrow 0$, we obtain
\beq
	\lim_{z_B\rightarrow 0}z_B^{-\Delta}G_F(x,x_B)=z^{\Delta_-}\delta^{d-1}(x-x_B)+\ldots \,,
\eeq
which is the correct boundary condition for a bulk to boundary propagator (\ref{eq:bbcond}).

Finally, we should show that (\ref{eq:claim1}) and (\ref{eq:claim2}) satisfy the right boundary conditions at the "turning point"
of the Schwinger-Keldysh contour. Using the definition of the Feynman correlator
\beq
G_F(x,x_B)=\theta(t-t_B)G_+(x,x_B)+\theta(t_B-t)G_+(x_B,x),
\eeq
we learn that when $t>t_B$ we have the identity
\beq
G_F(x,x_B)=G_+(x,x_B).
\eeq
Thus, (\ref{eq:claim1}) and (\ref{eq:claim2}) satisfy the continuity conditions (\ref{eq:mid1}) and (\ref{eq:mid2})
at $t=t_m$, for any choice of $t_m>t_B$. 

Finally, by using the SvR dictionary (\ref{eq:svrdict1}) and  (\ref{eq:svrdict2}), we obtain the boundary correlators
\begin{align}
&\langle T \mathcal{O}(x_1)\mathcal{O}(x_2)\rangle =(2\nu)^2\lim_{z_1\rightarrow 0}\lim_{z_2\rightarrow 0}z_1^{-\Delta}z_2^{-\Delta} G_F(x_1,x_2),
\\
&\langle \mathcal{O}(x_1)\mathcal{O}(x_2)\rangle =(2 \nu)^2 \lim_{z_1\rightarrow 0}\lim_{z_2\rightarrow 0}z_1^{-\Delta}z_2^{-\Delta}  G_+(x_1,x_2),
\end{align}
which proves the equivalence of the two dictionaries.

\section{Two point functions in AdS$_2$-Vaidya}

In the rest of this paper, we study examples of correlation functions in non-equilibrium settings in AdS/CFT. We consider a simple class of states, where the system is prepared in the vacuum state and is then suddenly perturbed out of equilibrium. The perturbation backreacts on the geometry and generically leads to black hole formation in the bulk. A simple model of this process is provided by the Vaidya spacetime. In the following, our main interest is the AdS$_3$-Vaidya spacetime, which is dual to a 1+1 dimensional CFT. In this Section, we first work out the simpler example of the AdS$_2$-Vaidya collapse as a warm-up problem. This serves as an illustration of the calculational method we use in the AdS$_3$-Vaidya case, with the advantage that the results in the AdS$_2$-Vaidya case are analytic. The results of this Section have been obtained earlier in \cite{Ebrahim:2010ra} and another closely related calculation can be found is \cite{Lowe:2008ra}.

\subsection{AdS$_2$-Vaidya spacetime}

The AdS$_2$ metric is given by
\begin{equation}
	ds^2=\frac{1}{z^2}\left(-dv^2-2dv\, dz\right)=\frac{1}{z^2}\left(-dt^2+dz^2\right),\quad \text{where} \quad t=v+z.
\end{equation}
For future convenience, we introduced an ingoing null coordinate $v$. Thus, $v=v_0=\text{constant}$ is an ingoing null ray that starts from the boundary at time $t=v_0$. Gravity in 1+1 dimensions is in certain sense trivial. For example, a well-known fact is that there are no gravitational waves in dimensions smaller than four. However, there are black hole solutions even in 1+1 dimensions (see e.g.\ \cite{Spradlin:1999bn} for a more thorough discussion). The one we will be considering has the metric
\begin{equation}
	ds^2=\frac{1}{z^2}\left(-(1-z^2)dv^2-2 dv\,dz\right),
\end{equation}
where we use an ingoing null coordinate labelled by the same letter $v$. This solution has a
horizon at $z=1$ and a Hawking temperature $T_H=1/2\pi$. Due to triviality of gravitational dynamics, 
this spacetime is in fact locally identical to AdS$_2$. Indeed, we can perform the following change of variables
\begin{equation}
	\bar{z}=\frac{z}{1+\cosh{v}+z\sinh{v}}\quad\text{and}\quad\bar{v}=\tanh\frac{v}{2},
	\label{eq:change_of_var}
\end{equation}
which leads to
\begin{equation}
	ds^2=\frac{1}{z^2}\Big(-\left(1-z^2\right)dv^2-2 dv\,dz\Big)=\frac{1}{\bar{z}^2}\left(-d\bar{v}^2-2d\bar{v}\,d\bar{z}\right).
\end{equation}
It is important to note that the black hole coordinates only cover a part of the AdS$_2$ spacetime \cite{Spradlin:1999bn}, which one might call the Rindler wedge of AdS$_2$. Thus, the thermal nature of quantum fields in the black hole spacetime can be understood as a manifestation of the Unruh effect \cite{Unruh:1976db} in AdS$_2$. In particular, correlation functions of quantum fields in the AdS$_2$ vacuum state correspond to thermal correlation functions in the black hole coordinates, with temperature given by the Hawking temperature $T_H=1/2\pi$.

When 1+1 dimensional gravity is coupled to matter, one can find solutions that describe the collapse of matter into a black hole. 
Perhaps the simplest case is the AdS$_2$-Vaidya spacetime
\begin{equation}
	ds^2=\frac{1}{z^2}\Big(-(1-\theta(v)z^2)dv^2-2dv\, dz\Big).
	\label{eq:BHmetric}
\end{equation}
This metric is identical to AdS$_2$ when $v<0$ and to the black hole metric when $v>0$. The metric functions have a discontinuity at $v=0$, where the collapsing matter is located. Thus, the matter is falling along a null geodesic. 

Above, we have only given the 1+1 dimensional metrics of interest, while we have not even specified what the gravitational theory is. For explicit theories leading to the above spacetimes, we refer the reader to the recent work \cite{Almheiri:2014cka} and the references therein.

\subsection{Calculation of correlation functions}

Next, we turn to study correlation functions of quantum fields in the AdS$_2$-Vaidya spacetime. For simplicity, we will consider the correlation functions of a massless scalar field with the action (\ref{eq:scalaraction}), with $m=0$ and $d=2$. This action is invariant under Weyl transformations $g_{\mu\nu}\rightarrow \Omega(x)g_{\mu\nu}$, which simplifies the task of calculating the two point functions. 

The Feynman correlator for a massless scalar in AdS$_2$ in the above coordinates is given by
\begin{equation}\label{eq:AdS2vac}
	G_F^\text{AdS}(v_2,z_2;v_1,z_1)=-\frac{1}{4\pi}\log\left(\frac{-(v_2-v_1)^2-2(z_2-z_1)(v_2-v_1)+i\epsilon}{-(v_2-v_1)^2-2(z_2-z_1)(v_2-v_1)+4z_2 z_1+i\epsilon}\right).
\end{equation}
Its form can be easily understood as follows. Since AdS$_2$ can be transformed to the upper half plane of Minkowski space (imagining that time runs horizontally) by a Weyl transformation, the two point function in AdS$_2$ can be calculated in Minkowski space with a Dirichlet boundary condition at $z=0$. The correlator of a massless scalar in 1+1 dimensions is given by $-\log(x^2+i\epsilon)/4\pi$. To satisfy Dirichlet boundary conditions at $z=0$, one must add a mirror contribution to the two point function from the mirror point at $(z_m,v_m)=(-z,v+2z)$.\footnote{In the $(z,t)$ coordinate system, the mirror point is at $(z_m,t_m)=(-z,t)$, which after transforming to the lightcone coordinates, becomes the formula presented in the main text.} These two together, give the two point function (\ref{eq:AdS2vac}). For another derivation of this result see e.g.\ \cite{Spradlin:1999bn}.

Given the coordinate transformation \eqref{eq:change_of_var} from the black hole metric to AdS$_2$, the Feynman correlator on the black hole side of the shell is
\begin{equation}
	G_F^\text{BH}(v_2,z_2;v_1,z_1)=G_F^\text{AdS}(\bar{v}_2,\bar{z}_2;\bar{v}_1,\bar{z}_1).
\end{equation}
As we explained, this is the two point function in a thermal state in the black hole spacetime, with temperature $T_H=1/2\pi$. In the following, we will use the barred coordinates in the $v>0$ part of the Vaidya spacetime. This is merely to simplify the algebra. The same result can be obtained by working in the original $(z,v)$ coordinates in the whole Vaidya spacetime. At $v=0$, the AdS$_2$ coordinates are matched to the barred coordinates as follows
\beq
z=\frac{\bar{z}}{2},\quad \bar{v}=0=v.
\eeq
As initial data, we know the two point functions in the $v<0$ part of the AdS-Vaidya spacetime. As discussed in Section \ref{sec:initialvalue}, the two point function from the AdS to the BTZ side of the AdS-Vaidya spacetime can be obtained by using the joining formula (\ref{eq:joiningform1}),
which in this case reads
\beq
	G^a_F(v_1,z_1;v_0,z_0) = i\int dz\,\sqrt{-g}g^{vz}\left(G_F^\text{AdS}(0,z;v_0,z_0)\overleftrightarrow{\partial_z}G_R^\text{BH}(v_1,z_1;0,z)\right).
\label{eq:joiningads2}
\eeq
where the point $(z_0,v_0)$ is in the AdS part, $v_0<0$, and the point $(z_1,v_1)$ is in the black hole part of the AdS-Vaidya spacetime, $v_1>0$. The superscript $a$ in (\ref{eq:joiningads2}) refers to the fact that this correlator is between points across the shell. Integrating (\ref{eq:joiningads2}) by parts we obtain a more convenient form
\beq
	G^a_F(v_1,z_1;v_0,z_0)= -2i\int dz\,G_F^\text{AdS}(0,z;v_0,z_0)\partial_z G_R^\text{BH}(v_1,z_1;0,z)
\eeq
where the boundary term vanishes, since
\begin{equation}
	G_R^\text{BH}(v_1,z_1;0,0)=0 \quad\text{and}\quad G_R^\text{BH}(v_1,z_1;0,1)=0.
\end{equation}
The retarded correlator can be determined from the Feynman correlator using the following identity
\begin{align}
	G_R^\text{BH}(v_1,z_1;0,z) &= 2i \Im{G_F^\text{AdS}(\bar{v}_1,\bar{z}_1;0,z/2)}\notag\\
		&= \frac{i}{2\pi}\text{Im}\left[\log(\bar{v}_1^2+(2\bar{z}_1+z)\bar{v}_1-2\bar{z}_1 z-i\epsilon)-\log(\bar{v}_1^2+(2\bar{z}_1+z)\bar{v}_1-i\epsilon)\right]\notag\\
		&= \frac{i}{2}\left[\theta\left(z-2\bar{z}_1-\bar{v}_1\right)-\theta\left(z-\bar{v}_1\right)\right],
\end{align}\
and therefore
\begin{equation}
	\partial_z G_R^\text{BH}(v_1,z_1;0,z) = \frac{i}{2}\left[\delta\left(z-2\bar{z}_1-\bar{v}_1\right)-\delta\left(z-\bar{v}_1\right)\right].
\end{equation}
We can evaluate the above integral to obtain
\begin{align}
	G^a_F(v_1,z_1;v_0,z_0) &=\int dz\,G_F^\text{AdS}(0,z;v_0,z_0)\left[\delta\left(z-2\bar{z}_1-\bar{v}_1\right)-\delta\left(z-\bar{v}_1\right)\right]\notag\\
		&= G_F^\text{AdS}(0,2\bar{z}_1+\bar{v}_1;v_0,z_0)-G_F^\text{AdS}(0,\bar{v}_1;v_0,z_0).
	\label{eq:Gfacross}
\end{align}

In the next step, we use the fact that the correlator $G_F^a$ also satisfies the Klein--Gordon equation of motion (with a delta function source) with respect to the variable $(z_0,v_0)$. Thus, we can propagate that point past the shell to the black hole spacetime by using the joining formula (\ref{eq:joining3propa}), visualized in Figure \ref{fig:joining2},
\begin{equation}
	G_F(v_2,z_2;v_1,z_1)=-2i\int dz_0\, G^a_F(v_2,z_2;v_0,z_0)\partial_{z_0}G^a_R(v_1,z_1;v_0,z_0).
\end{equation}
We will again only obtain $\delta$-function contributions by first observing
\begin{align}
	G_R^a(v_1,z_1;v_0,z_0) &= 2i \,\text{Im}\, G_F^a(v_1,z_1;v_0,z_0)\notag\\
		&=2i \,\text{Im}\left[G_F^\text{AdS}(0,2\bar{z}_1+\bar{v}_1;v_0,z_0)-G_F^\text{AdS}(0,\bar{v}_1;v_0,z_0)\right]\notag\\
		&=\frac{i}{2}\left[\theta\left(z_0-2\bar{z}_1-\bar{v}_1+\frac{v_0}{2}\right)-\theta\left(z_0+\frac{v_0}{2}\right)-\theta\left(z_0-\bar{v}_1+\frac{v_0}{2}\right)+\theta\left(z_0+\frac{v_0}{2}\right)\right]\notag\\
	&=\frac{i}{2}\left[\theta\left(z_0-2\bar{z}_1-\bar{v}_1+\frac{v_0}{2}\right)-\theta\left(z_0-\bar{v}_1+\frac{v_0}{2}\right)\right],
\end{align}
and therefore
\begin{equation}
	\partial_{z_0} G_R^a(z_1,v_1,z_0,v_0) = \frac{i}{2}\left[\delta\left(z_0-2\bar{z}_1-\bar{v}_1+\frac{v_0}{2}\right)-\delta\left(z_0-\bar{v}_1+\frac{v_0}{2}\right)\right].
\end{equation}
Evaluating the $z_0$-integral and using \eqref{eq:Gfacross}, we find
\begin{align}
	G_F(v_2,z_2;v_1,z_1) = & \int dz_0\, G_F^a(v_2,z_2;v_0,z_0)\left[\delta\left(z_0-2\bar{z}_1-\bar{v}_1+\frac{v_0}{2}\right)-\delta\left(z_0-\bar{v}_1+\frac{v_0}{2}\right)\right]\notag\\
		= &\; G_F^a\left(v_2,z_2;v_0,2\bar{z}_1+\bar{v}_1-\frac{v_0}{2}\right)-G_F^a\left(v_2,z_2;v_0,\bar{v}_1-\frac{v_0}{2}\right)\notag\\
		= &\; G_F^\text{AdS}\left(0,2\bar{z}_2+\bar{v}_2;v_0,2\bar{z}_1+\bar{v}_1-\frac{v_0}{2}\right)-G_F^\text{AdS}\left(0,\bar{v}_2;v_0,2\bar{z}_1+\bar{v}_1-\frac{v_0}{2}\right)\notag\\
		&-G_F^\text{AdS}\left(0,2\bar{z}_2+\bar{v}_2;v_0,\bar{v}_1-\frac{v_0}{2}\right)+G_F^\text{AdS}\left(0,\bar{v}_2;v_0,\bar{v}_1-\frac{v_0}{2}\right)\notag\\
		= &\; \frac{1}{4\pi}\log\left(\frac{(\bar{v}_2-\bar{v}_1)^2+2(\bar{z}_2-\bar{z}_1)(\bar{v}_2-\bar{v}_1)-4\bar{z}_2 \bar{z}_1-i\epsilon}{(\bar{v}_2-\bar{v}_1)^2+2(\bar{z}_2-\bar{z}_1)(\bar{v}_2-\bar{v}_1)-i\epsilon}\right)\notag\\
		= &\; G_F^{\text{BH}}(v_2,z_2;v_1,z_1).
\end{align}
We can therefore conclude that the bulk Feynman correlator becomes instantly thermal when both points of the correlation function are outside of the infalling shock wave, i.e. for $v_1>0$ and $v_2>0$. This result was indeed obtained in \cite{Ebrahim:2010ra}. We can also define boundary correlation functions using the "extrapolate" dictionary leading to
\begin{align}
	G_F(t_2,t_1)=&\frac{\theta(-t_1)\theta(-t_2)}{-(t_2-t_1)^2+i\epsilon}+\frac{\theta(t_2)\theta(-t_1)}{-(2\sinh \frac{t_2}{2}-t_1 \cosh \frac{t_2}{2})^2+i\epsilon}\nonumber\\
		&+\frac{\theta(t_1)\theta(-t_2)}{-(2\sinh \frac{t_1}{2}-t_2 \cosh\frac{t_1}{2})^2+i\epsilon}+\frac{\theta(t_1)\theta(t_2)}{2(1-\cosh(t_2-t_1))-i\epsilon}
\end{align}
One interesting thing to note about the result is that it agrees exactly with the results obtained in \cite{Balasubramanian:2012tu} using the 
(complex) geodesic approximation and after setting $\Delta=1$. While the calculation in \cite{Balasubramanian:2012tu} was obtained in the context of the AdS$_3$-Vaidya spacetime, the same geodesics can be used in AdS$_2$-Vaidya leading to the same result.

\section{Two point functions in AdS$_3$-Vaidya spacetime}

\subsection{AdS$_3$-Vaidya spacetime}
	\label{sec:ads3_vaidya}

Gravity in 2+1 dimensions is trivial in a similar sense as 1+1 dimensional gravity. There are no local graviton excitations and the dynamics of spacetime is directly related to the dynamics of the matter in the theory. 2+1 dimensional gravity still has black hole solutions, which makes it an interesting system to study. The black hole we will be studying is the BTZ black hole \cite{Banados:1992wn} (or more precisely black string) with the metric
\begin{equation}
	ds_\text{BTZ}^2=\frac{1}{z^2}\left[-(1-z^2)dt^2+\frac{dz^2}{1-z^2}+dx^2\right].
	\label{eq:btz_metric}
\end{equation}
The BTZ black hole has an event horizon at $z=1$ and a Hawking temperature $T_H=1/2\pi$.
As in the case of 1+1 gravity, there is a coordinate transformation that maps the BTZ metric to the AdS$_3$
metric
\begin{align}
	\bar{x} &=\sqrt{1-z^2}e^{x}\cosh(t),\notag\\
	\bar{t} &=\sqrt{1-z^2}e^{x}\sinh(t),\notag\\
	\bar{z} &=ze^{x},
	\label{eq:coordchange}
\end{align}
with
\begin{equation}
	ds_\text{BTZ}^2=\frac{1}{\bar{z}^2}(-d\bar{t}^2+d\bar{z}^2+d\bar{x}^2).
\end{equation}
A crucial feature of the coordinate transformation (\ref{eq:coordchange}) is that the black hole coordinates again cover only a part of the AdS$_3$ spacetime. This part is often called, the AdS-Rindler wedge (see e.g.\ \cite{Bousso:2012mh}). As in the AdS$_2$ case, quantum fields in the vacuum state in AdS$_3$ will be in a thermal state in the BTZ spacetime, due to the Unruh effect.

The BTZ black hole can be formed by collapse of matter. In the following, we will be studying the 
AdS$_3$-Vaidya spacetime that corresponds to the collapse of null shock wave starting from the boundary
at time $v=0$.
The metric of the AdS$_3$-Vaidya spacetime is given by
\begin{equation}
	ds^2=\frac{1}{z^2}\left[-(1-\theta(v)z^2)dv^2-2dv\, dz+dx^2\right].
	\label{eq:vaidyar}
\end{equation}
It is easy to see that the metric (\ref{eq:vaidyar}) is identical to the AdS$_3$ metric with $v=t-z$ for $v<0$. For $v>0$, the metric (\ref{eq:vaidyar}) is identical to the BTZ metric (\ref{eq:btz_metric}) with $v=t-\frac{1}{2}\log((1-z)/(1+z))$.

\subsection{Correlation functions in AdS$_3$ and BTZ backgrounds}

Again, we will consider a free scalar field with the action (\ref{eq:scalaraction}) with $m^2=-3/4$ and $d=3$. In the boundary theory $\phi$ is dual to a scalar operator $\mathcal{O}$ with scaling dimension $\Delta=3/2$. We have chosen the value of the mass $m^2=-3/4$ since, with this choice, the action (\ref{eq:scalaraction}) can be written in a form invariant under Weyl transformations $g_{\mu\nu}\rightarrow \Omega^2(x)g_{\mu\nu}$ and $\phi\rightarrow \Omega^{-1/2}(x)\phi$ (see e.g.\ \cite{Birrell:1982ix}). This symmetry simplifies the form of the bulk correlation functions.

The Feynman correlator for a scalar field with scaling dimension $\Delta=3/2$ in AdS$_3$ is
\begin{align}
	G_{F}^{\text{AdS}}(v_2,x_2,z_2&;v_1,x_1,z_1)\notag\\
		&=\frac{\sqrt{z_1 z_2}}{4\pi}\left(\frac{1}{\sqrt{-(v_2-v_1)^2-2(v_2-v_1)(z_2-z_1)+(x_2-x_1)^2+i\epsilon}}\right.\notag\\
		&\qquad\qquad \left. -\frac{1}{\sqrt{-(v_2-v_1)^2-2(v_2-v_1)(z_2-z_1)+4z_1 z_2+(x_2-x_1)^2+i\epsilon}}\right).
	\label{eq:GfAdS_xspace}
\end{align}
The result can be understood as follows. By a Weyl transformation with $\Omega=z$, one can transform AdS$_3$ into
the upper half-plane of Minkowski spacetime. The transformation takes the equation of motion of $\phi$ into the equation of motion of a massless scalar. The two point function of a massless scalar in Minkowski space is $1/(4\pi\sqrt{x^2+i\epsilon})$. To satisfy Dirichlet boundary conditions at $z=0$, one must include a contribution from a mirror point at $(z_m,v_m)=(-z,v+2z)$. Combining these two leads to the correlation function (\ref{eq:GfAdS_xspace}), where the overall factor of $\sqrt{z_1z_2}$ appears from the Weyl transformation of the operator $\phi$ when one transforms back to AdS$_3$. Another derivation of (\ref{eq:GfAdS_xspace}) can be found in \cite{Satoh:2002bc}.

Translational invariance along the $x$-direction allows us to Fourier transform the two point function,
\begin{align}
	G_{F}^{\text{AdS}}&(v_2,z_2;v_1,z_1;k) = \int dx\, G_{F}^{\text{AdS}}(v_2,x,z_2;v_1,0,z_1)e^{-ikx}\notag\\
		&=\frac{\sqrt{z_1 z_2}}{2\pi}\left[K_0\left(\sqrt{-(v_2-v_1)^2-2(v_2-v_1)(z_2-z_1)+i\epsilon}\,|k|\right)\right.\notag\\
		&\qquad\qquad\quad \left. -K_0\left(\sqrt{-(v_2-v_1)^2-2(v_2-v_1)(z_2-z_1)+4z_1 z_2+i\epsilon}\,|k|\right)\right],
	\label{eq:GfAdS}
\end{align}
where $K_0$ is a modified Bessel function. 

\begingroup
\allowdisplaybreaks

As we argued earlier, the ground state in AdS$_3$ gets mapped in the coordinate transformation (\ref{eq:coordchange}) to a thermal state with the temperature given by the Hawking temperature $T_H=1/2\pi$. Thus, the two point function in a thermal state in the BTZ spacetime can be obtained by a coordinate transformation of the AdS$_3$ two point function\footnote{As a check, one can easily verify that (\ref{eq:GfBTZaux}) is periodic in imaginary time, with a period $2\pi$.} (\ref{eq:GfAdS_xspace})
\begin{align}
	&G_{F}^{\text{BTZ}}(v_2,x_2,z_2;v_1,x_1,z_1) = G_{F}^{\text{AdS}}(\bar{v}_2,\bar{x}_2,\bar{z}_2;\bar{v}_1,\bar{x}_1,\bar{z}_1)\label{eq:GfBTZaux}\\
		&=\sqrt{\frac{z_1 z_2}{32\pi^2}}\left(\frac{1}{\sqrt{\cosh(x_2-x_1)-z_1 z_2-(1-z_1 z_2)\cosh(v_2-v_1)-(z_2-z_1)\sinh(v_2-v_1)+i\epsilon}}\right.\notag\\
		&\qquad\qquad\quad \left. -\frac{1}{\sqrt{\cosh(x_2-x_1)+z_1 z_2-(1-z_1 z_2)\cosh(v_2-v_1)-(z_2-z_1)\sinh(v_2-v_1)+i\epsilon}}\right)\notag	
\end{align}
To compute the Fourier transform of the BTZ-correlator, we use the following integral identity
\begin{align}
	\int dx\,\frac{e^{-ikx}}{\cosh(x)^p} &= \int dx\, e^{-ikx}\left(\frac{e^x+e^{-x}}{2}\right)^{-p}\notag\\
		&= 2^p \int dx\, e^{-(ik+p)x}\left(1+e^{-2x}\right)^{-p}\notag\\
		&= \frac{2^p}{\Gamma(p)}\int_0^\infty du\,\int dx\, e^{-(ik+p)x}u^{p-1}e^{-u(1+e^{-2x})}\notag\\
		&=\frac{2^{p-1}}{\Gamma(p)}\int_0^\infty du\, u^{\frac{p-ik}{2}-1}e^{-u}\int_0^\infty dy\, y^{\frac{p+ik}{2}-1}e^{-y}\qquad\qquad y(x)=u e^{-2x}\notag\\
		&=\frac{2^{p-1}}{\Gamma(p)}\Gamma\left(\frac{p-ik}{2}\right)\Gamma\left(\frac{p+ik}{2}\right),
\end{align}
which gives
\begin{align}
	\int dx\, \frac{e^{-ikx}}{\sqrt{\cosh(x)+b}} &= \int dx\, \frac{e^{-ikx}}{\sqrt{\cosh(x)}}\frac{1}{\sqrt{1+\frac{b}{\cosh(x)}}}\notag\\
		&= \int dx\, \frac{e^{-ikx}}{\sqrt{\cosh(x)}}\sum_{n=0}^\infty \binom{-\frac{1}{2}}{n}\frac{b^n}{\cosh(x)^n}\notag\\
		&= \sum_{n=0}^\infty\frac{\Gamma\left(\frac{1}{2}\right)b^n}{\Gamma\left(n+1\right)\Gamma\left(\frac{1}{2}-n\right)}\int dx\,\frac{e^{-ikx}}{\cosh(x)^{n+\frac{1}{2}}}\notag\\
		&=\sum_{n=0}^\infty\frac{\Gamma\left(\frac{1}{2}\right)b^n}{\Gamma\left(n+1\right)\Gamma\left(\frac{1}{2}-n\right)}\frac{2^{n-\frac{1}{2}}}{\Gamma\left(\frac{1}{2}+n\right)}\Gamma\left(\frac{n+\frac{1}{2}+ik}{2}\right)\Gamma\left(\frac{n+\frac{1}{2}-ik}{2}\right)\notag\\
		&= \sum_{n=0}^\infty\frac{(-1)^n(2b)^n}{\sqrt{2\pi}n!}\Gamma\left(\frac{n+\frac{1}{2}+ik}{2}\right)\Gamma\left(\frac{n+\frac{1}{2}-ik}{2}\right)\notag\\
		&= \frac{1}{\sqrt{2\pi}}\left[\Gamma\left(\frac{1}{4}-\frac{ik}{2}\right)\Gamma\left(\frac{1}{4}+\frac{ik}{2}\right){}_2F_1\left(\frac{1}{4}-\frac{ik}{2},\frac{1}{4}+\frac{ik}{2},\frac{1}{2},b^2\right)\right.\notag\\
		&\qquad\qquad \left.-2b\Gamma\left(\frac{3}{4}-\frac{ik}{2}\right)\Gamma\left(\frac{3}{4}+\frac{ik}{2}\right){}_2F_1\left(\frac{3}{4}-\frac{ik}{2},\frac{3}{4}+\frac{ik}{2},\frac{3}{2},b^2\right)\right],
\end{align}
We can use this to get the Fourier transform of \eqref{eq:GfBTZaux}:
\begin{align}
	G_{F}^{\text{BTZ}}&(v_2,z_2;v_1,z_1;k)=\int dx\,e^{-ikx} G_{F}^{\text{BTZ}}(v_2,x,z_2;v_1,0,z_1)\notag\\
		&=\frac{1}{4\pi}\sqrt{\frac{z_1 z_2}{4\pi}}\left[\Gamma\left(\frac{1}{4}-\frac{ik}{2}\right)\Gamma\left(\frac{1}{4}+\frac{ik}{2}\right){}_2F_1\left(\frac{1}{4}-\frac{ik}{2},\frac{1}{4}+\frac{ik}{2},\frac{1}{2},b_1^2\right)\right.\notag\\
		&\qquad\qquad\qquad\quad -2b_1\Gamma\left(\frac{3}{4}-\frac{ik}{2}\right)\Gamma\left(\frac{3}{4}+\frac{ik}{2}\right){}_2F_1\left(\frac{3}{4}-\frac{ik}{2},\frac{3}{4}+\frac{ik}{2},\frac{3}{2},b_1^2\right)\notag\\
		&\qquad\qquad\qquad\quad -\Gamma\left(\frac{1}{4}-\frac{ik}{2}\right)\Gamma\left(\frac{1}{4}+\frac{ik}{2}\right){}_2F_1\left(\frac{1}{4}-\frac{ik}{2},\frac{1}{4}+\frac{ik}{2},\frac{1}{2},b_2^2\right)\notag\\
		&\qquad\qquad\qquad\quad\left. +2b_2\Gamma\left(\frac{3}{4}-\frac{ik}{2}\right)\Gamma\left(\frac{3}{4}+\frac{ik}{2}\right){}_2F_1\left(\frac{3}{4}-\frac{ik}{2},\frac{3}{4}+\frac{ik}{2},\frac{3}{2},b_2^2\right)\right],
	\label{eq:GfBTZ}
\end{align}
where
\begin{align}
	b_1=b_1(v_2,z_2;v_1,z_1) &= -z_1 z_2-(1-z_1 z_2)\cosh(v_2-v_1)-(z_2-z_1)\sinh(v_2-v_1)+i\epsilon,\notag\\
	b_2=b_2(v_2,z_2;v_1,z_1) &= z_1 z_2-(1-z_1 z_2)\cosh(v_2-v_1)-(z_2-z_1)\sinh(v_2-v_1)+i\epsilon.
	\label{eq:b1_and_b2}
\end{align}
Finally, the retarded correlator in the BTZ spacetime is given by 
\beq
G_R(v_2,z_2;v_1,z_1;k)=2i\,\theta(v_2-v_1)\,\textrm{Im}\,G_{F}^{\text{BTZ}}(v_2,z_2;v_1,z_1;k),\label{eq:retardedBTZ}
\eeq
where we replaced $\theta(t_2-t_1)$ by $\theta(v_2-v_1)$ which can be justified by $\text{Im}\,G_F$ vanishing for spacelike separated points.
\endgroup

\subsection{Calculation of the AdS$_3$-Vaidya correlator}

As we know the ground state correlator in AdS$_3$ and the retarded correlator (\ref{eq:retardedBTZ}) in the BTZ spacetime, we are ready to use the joining formula (\ref{eq:joining3propa}) to obtain the two point function in the full AdS$_3$-Vaidya spacetime. For the purpose of numerics, (\ref{eq:joining3propa}) is somewhat problematic. First of all, there are six numerical integrations one should perform, three over the $z$-coordinates of the joining points and three over the corresponding $x$-coordinates. Thus, we have a six dimensional integral to perform. Another problem is that the two point functions that one should integrate over have branch point singularities at the lightcone. These singularities are of the type $1/\sqrt{s^2}$ and $1/(s^2)^{3/2}$, where $s^2$ is the square of a Lorentzian distance. Thus, a direct numerical integration over the lightcone seems difficult. However, we can use $x$ translational invariance to Fourier transform the joining formula (\ref{eq:joining3propa}), which in momentum space becomes
\begin{align}
	G_F(v_4,z_4;v_3,z_3;k)=-i \int dz_2 dz_1& dz_0 \Big(G^\text{AdS}_F(0,z_1;v_0,z_0;k)\overleftrightarrow{D}^{v_1}G^\text{BTZ}_R(v_4,z_4;0,z_1;k)\Big)\times\notag\\
		&\times\overleftrightarrow{D}^{v_0}\Big(G^\text{AdS}_R(0,z_2;v_0,z_0;k) \overleftrightarrow{D}^{v_2}G_R^\text{BTZ}(v_3,z_3;0,z_2;k)\Big),\label{eq:joining3propak}
\end{align}
where the two point functions appearing in the integrand are given by (\ref{eq:GfAdS}) and (\ref{eq:GfBTZ}). This form (\ref{eq:joining3propak}) has two advantages as compared to the position space integral (\ref{eq:joining3propa}). Firstly, it has only three integrations instead of six. Secondly, the lightcone divergences in the momentum space two point functions are only logarithmic $\propto\log s$. Integrals over logarithms can be straightforwardly integrated numerically. We have found it useful to use Mathematica's NIntegrate, together with the command Exclusions, which allows us to exclude the singular points from the numerical integrals. Then, the singular points have to be dealt with analytically. We describe the details of the procedure in Appendix \ref{sec:details}. The numerical accuracy of our methods is studied in Appendix \ref{sec:numerics}. Next, we move on to discuss the results of the numerical integration of (\ref{eq:joining3propak}).

\subsection{Thermalization of the Feynman correlator}
\label{sec:Feynman_thermalization}

\begin{figure}
\centering
\includegraphics[width=15cm]{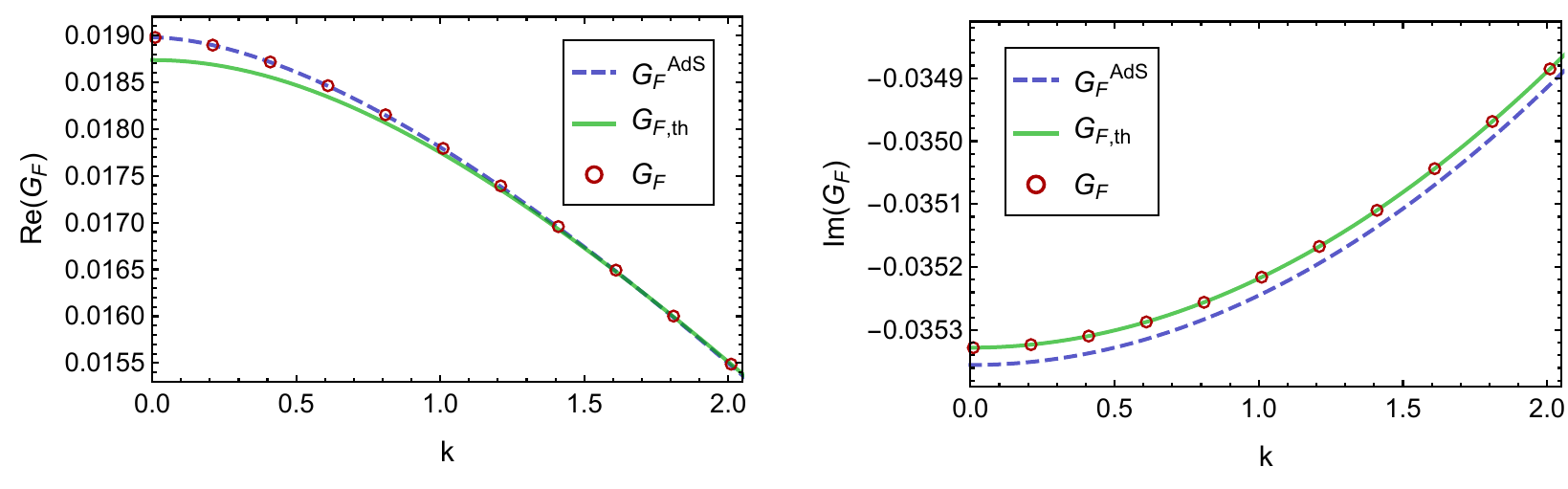}
\caption{\label{fig:thermalization1} The real (left) and imaginary (right) parts of the bulk Feynman correlator at lightcone times $v_2=0.051$ and $v_1=0.001$ right after the collapse and for $z_1=0.1$ and $z_2=0.2$, as a function of momentum. The blue curves are AdS vacuum correlators while the green curves are thermal BTZ correlators. The red dots are our results for the Vaidya correlator. We can see that the real part of the Vaidya correlator agrees with the AdS one right after the collapse while the imaginary part thermalizes immediately.}
\end{figure}

\begin{figure}
	\centering
	\includegraphics[width=13.5cm]{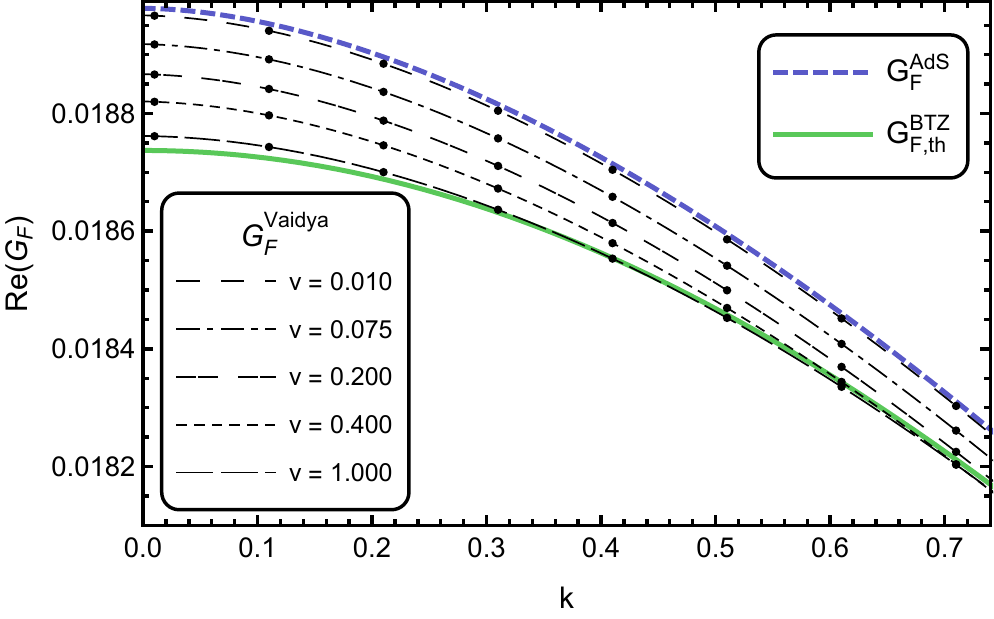}
	\caption{The real part of the bulk Feynman correlator $G_{F}(v_2,z_2;v_1,z_1;k)$ in AdS$_3$ (blue), BTZ (green) and Vaidya (black, dashed) as a function of momentum. We choose the parameters $z_1=0.1$, $z_2=0.2$ and $\d v=v_2-v_1=0.05$. For the Vaidya curves, $v_1=\{0.010,0.075,0.20,0.40,1.00\}$ with $v_1$ growing from top to bottom. We can see that the correlator is close to the AdS value initially before it approaches the thermal BTZ correlator at late times.}
	\label{fig:thermalization2}
\end{figure}

Using \eqref{eq:joining3propak}, we can now compute the Feynman correlator on the BTZ side of the Vaidya geometry. Firstly, we see that the imaginary part of the propagator thermalizes instantly after the quench, as shown on the right hand side of Figure \ref{fig:thermalization1} for the bulk to bulk correlator. This is expected as the imaginary part of the Feynman propagator is related to the retarded correlator, which does not depend on the quantum state for a free field. Therefore, the imaginary part is not affected by the excitation due to the collapse and immediately settles for the BTZ thermal state value. The real part of the Feynman correlator approaches its thermal value at different rates depending on the value of the momentum. On the left hand side of Figure \ref{fig:thermalization1}, one can see that the real part still equals the AdS value right after the collapse, which is in great contrast to the imaginary part. As time evolves the imaginary part stays constant at the thermal value. Moreover, the real part approaches the thermal value as can be seen from Figure \ref{fig:thermalization2}. There, we plot the real part of the Vaidya correlator for different values of the lightcone time $v_1$ after the shock wave while keeping $\delta v=v_2-v_1$ constant. We can see that the real part of the Vaidya correlator, being close to the AdS correlator right behind the shock wave, approaches the thermal correlator at later times.

\begin{figure}
	\centering
	\includegraphics[width=13.5cm]{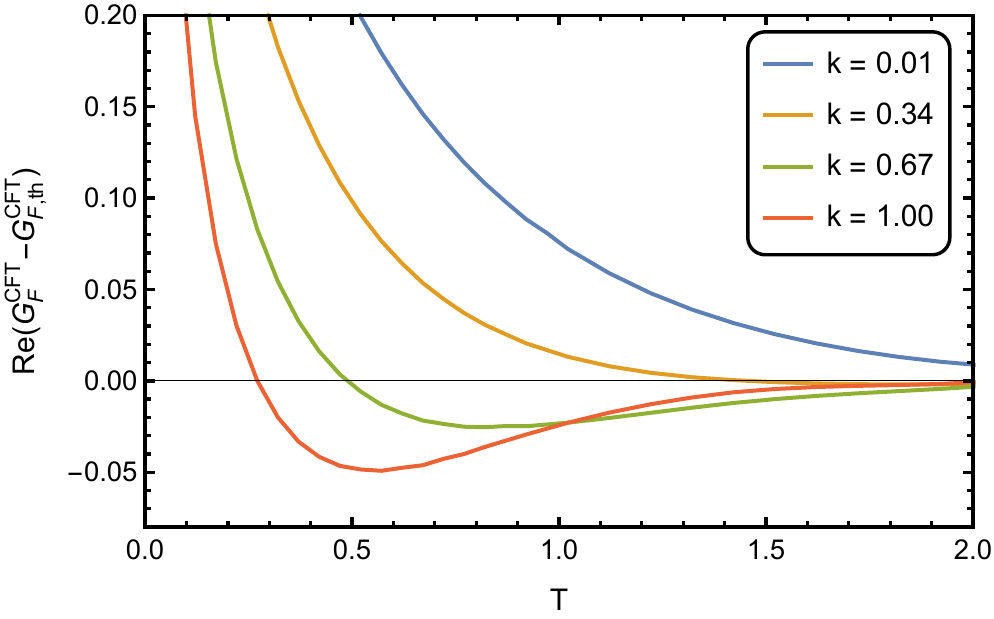}
	\caption{The difference of the real part of the boundary Feynman correlator (\ref{eq:CFT_correlator}) from the thermal correlator as a function of the average time $T=\frac{t_2+t_1}{2}$. The evolution is plotted for several different momentum modes with $k$ increasing from the top to the bottom curve. The parameters are $\epsilon_1=\epsilon_2=0.001$ and $\d t=t_2-t_1=0.02$.}
	\label{fig:thermalization3}
\end{figure}

In Figure \ref{fig:thermalization3}, we plot the difference of the real part of the boundary Feynman correlator from its thermal value. We consider correlators between points very close to the boundary and rescale the correlators according to the "extrapolate" dictionary (\ref{eq:2pointextrapolate}) to obtain the boundary field theory correlator even though we cannot take the strict boundary limit. Thus, we work with the finite cutoff version
\beq
	G_F^\text{CFT}(x_2,x_1)=2\pi\epsilon_1^{-\Delta}\epsilon_2^{-\Delta}G_F(x_2,\epsilon_2;x_1,\epsilon_1),
	\label{eq:CFT_correlator}
\eeq
where the overall factor of $2\pi$ is chosen so that the vacuum two point function has a canonical normalization $1/|x_1-x_2|^{2\Delta}$. The different curves correspond to different momenta, while $T$ is the average boundary time
\beq
	T=\frac{t_1+t_2}{2}.
\eeq
We see that the two point function approaches the thermal one at a different rate for different values of the momentum.
The slowest approach is found for the smallest value of the momentum $k=0.01$. Defining a strict thermalization time for the correlation function seems difficult, as the correlator is found to oscillate around the thermal value. The higher momenta fall of more rapidly at early times and then undershoot the thermal value and oscillate around it. 

\begin{figure}
	\centering
	\includegraphics[width=13.5cm]{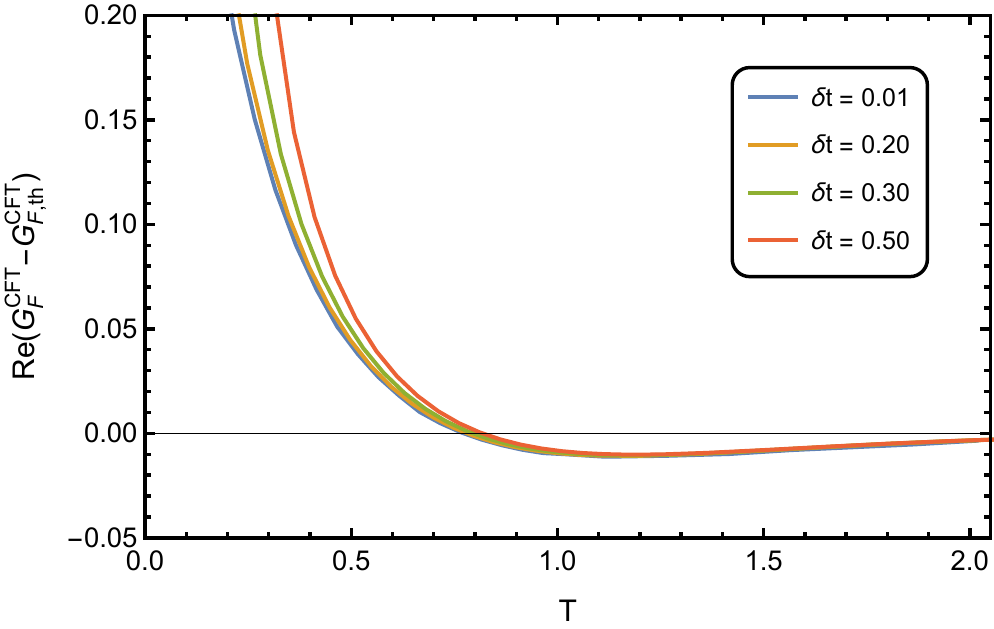}
	\caption{The difference of the real part of the boundary Feynman correlator (\ref{eq:CFT_correlator}) from the thermal correlator as a function of the average time $T=\frac{t_2+t_1}{2}$. The different curves now correspond to different values of the time difference $\delta t=t_2-t_1$, with $\delta t$ increasing from the bottom curve towards the top curve. The value of the momentum is $k=0.5$ and the other parameters are $\epsilon_1=\epsilon_2=0.001$.}
	\label{fig:delta_t}
\end{figure}

One can also ask how the approach to the thermal limit depends on the time difference $\delta t=t_2-t_1$ in the boundary correlation function. Again, the imaginary part is thermal as long as both of the points are at positive times. The time evolution of the Feynman correlator is shown in Figure \ref{fig:delta_t} as a function of the average time $T$, while the different curves correspond to different values of $\delta t$. Figure \ref{fig:delta_t} shows that the time evolution of the correlator is fairly insensitive to the value of $\delta t$ at sufficiently late times, as all the different curves essentially collapse to one curve. In contrast, the early time dependence, when $\delta t$ is close to $T$, is sensitive to the value of $\delta t$.

\subsection{Why does the Feynman correlator thermalize?}

We have seen in the previous Section \ref{sec:Feynman_thermalization} that the Feynman correlator in AdS$_3$-Vaidya spacetime approaches the thermal correlator soon after the collapse. Viewed as a free bulk quantum field in a time dependent spacetime, it might at first seem surprising that the correlator approaches the thermal value at all as free field theories generically do not thermalize.\footnote{One way of seeing this is to note that free field theories have an infinite number of conserved quantum numbers, the occupation numbers $a^{\dagger}_k a_k$ for each value of $k$.} Thermalization of free fields in a collapsing black hole spacetime at very late times was explained by Hawking in \cite{Hawking:1974sw}. In this Section, we will see how thermalization appears using a formalism more similar to the one in the current paper. Closely related discussions have appeared e.g.\ in \cite{CaronHuot:2011dr}.

\begin{figure}
	\centering
	\includegraphics[width=10cm]{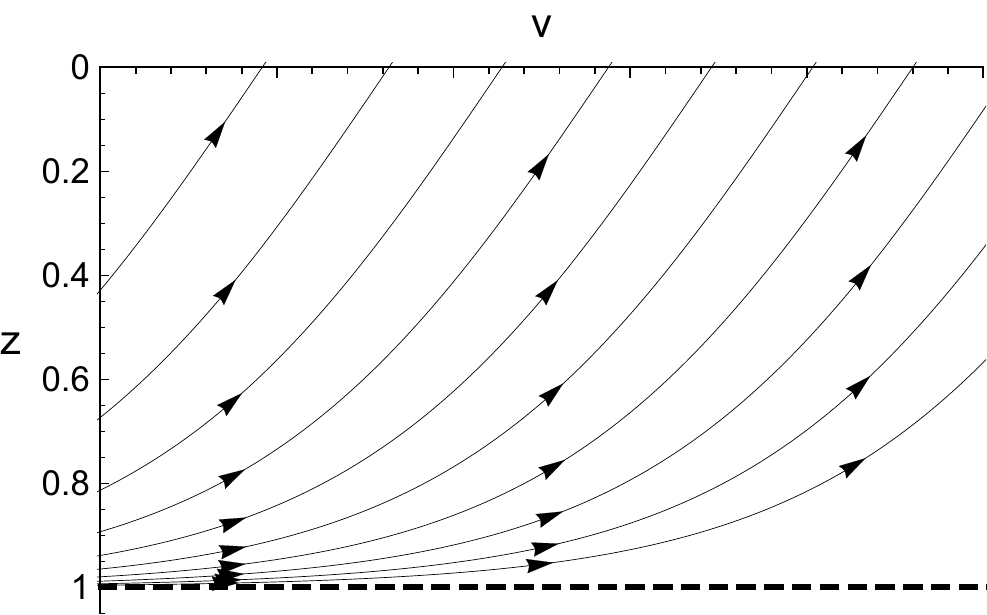}
	\caption{Congruence of outgoing null geodesics in Vaidya spacetime after the black hole has formed. The dashed line is the horizon. The initial radial position at $v=0$ approaches the horizon exponentially for the geodesics shown here. The closer the initial position is to the horizon, the longer it takes for the geodesic to reach the boundary.}
	\label{fig:geodesics}
\end{figure}

The retarded correlator propagates the initial data on the shell to the black hole part of spacetime. The retarded BTZ correlator between a point $(v_1=0,z_1)$ on the shell and a point outside the black hole $(v_2>0,z_2)$ falls off away from the lightcone as
\begin{equation}
	G_R(v_2,z_2;0,z_1;k)\propto e^{- v_2 \Delta}.
\end{equation}
As we are interested in the correlators at the late time limit $v_2\rightarrow\infty$, this contribution can be neglected. On the other hand, there are logarithmic divergences when $(0,z_1)$ and $(v_2,z_2)$ are separated by radial null geodesics, around which the retarded correlator is order one (as a function of $v_2$). Thus, in order to find the region of space at $v=0$ where the initial data is most relevant for late times, we should understand the relevant radial null geodesics. The ingoing null geodesics are simply lines of constant $v$.  On the other hand, the outgoing geodesics are given by
\begin{equation}
	z(v)=\frac{1+z_1-e^v(1-z_1)}{1+z_1+e^v(1-z_1)},
\end{equation}
where $z_1$ is the $z$-coordinate on the initial slice $v=0$. A congruence of outgoing null geodesics starting at $v=0$ is shown in Figure \ref{fig:geodesics}, where the initial radial position $z_1$ approaches the horizon exponentially for the plotted geodesics. Once the geodesics reach the boundary, they are reflected and fall into the horizon. Therefore, $(v_2,z_2)$ is reached by a direct and a reflected geodesic, which started out at the shell at $z_1=z_\pm(v_2,z_2)$, where $z_\pm$ are defined in (\ref{eq:z_plus_minus}). These two points on the shell, which are lightlike separated from $(v_2,z_2)$, approach the horizon exponentially for growing $v_2$ as
\begin{equation}
	z_-(v_2,z_2)= 1-\frac{2(1-z_2)}{1+z_2}e^{-v_2}+\mathcal{O}\left(e^{-2v_2}\right) \quad\text{and}\quad z_+(v_2,z_2)= 1-2e^{-2v_2}+\mathcal{O}\left(e^{-4v_2}\right).
\end{equation}
Thus, the part of the initial data that is relevant for the late time boundary two point function is located exponentially close to the horizon. As we will soon review, the near horizon correlator splits into ingoing and outgoing parts. The part relevant for the late time dynamics is the outgoing part only, as the ingoing part rapidly falls through the horizon into the black hole. Consequently, if we show that the outgoing part of the near-horizon Feynman correlator in Vaidya spacetime is thermal after the collapse, then we expect the full correlator in Vaidya spacetime to thermalize at late times.

Our next task is to calculate the outgoing part of the near horizon correlator. To do this, we first note that for $v>0$ the Klein--Gordon equation reduces to the equation of motion of a massless 1+1 dimensional scalar near the horizon. This can be seen by using the coordinate
\begin{equation}
	z_*=\frac{1}{2}\log\left(\frac{1+z}{1-z}\right)\Leftrightarrow z=\tanh z_*,
	\label{eq:z_star}
\end{equation}
in terms of which the Klein--Gordon equation satisfied by the Feynman propagator becomes
\begin{equation}
	\left(\partial_{{z_1}_*}^2-\partial_{t_1}^2\right)G_F(x_2,x_1)+\mathcal{O}\left(e^{-2{z_1}_*}\right) = 0 = \left(\partial_{{z_2}_*}^2-\partial_{t_2}^2\right)G_F(x_2,x_1)+\mathcal{O}\left(e^{-2{z_2}_*}\right),
\end{equation}
where we use the Schwarzschild time coordinate $t$. Since $\partial_{{z_i}_*}^2-\partial_{t_i}^2 = \left(\partial_{{z_i}_*}-\partial_{t_i}\right)\left(\partial_{{z_i}_*}+\partial_{t_i}\right)$, the near-horizon propagator can be split into two parts, one of which only depends on ${z_i}_*-t_i=v_i$ and is therefore ingoing and the other one only depends on ${z_i}_*+t_i=v_i+2{z_i}_*$ and is therefore outgoing. This applies to both $i=1,2$ and we can therefore split the near-horizon Feynman two point function according to
\begin{align}
	G_F(x_2,x_1) =\:& G_{F,oo}(v_2+2{z_2}_*,v_1+2{z_1}_*)+G_{F,io}(v_2,v_1+2{z_1}_*)\notag\\
		& +G_{F,oi}(v_2+2{z_2}_*,v_1)+G_{F,ii}(v_2,v_1),
	\label{eq:propagator_decomposition}
\end{align}
where the indices $i$ and $o$ indicate ingoing and outgoing parts with respect to the two coordinates. In order to calculate the outgoing part $G_{F,oo}$ right after the shock wave, we need the initial data from the AdS vacuum correlator. The Fourier-transformed AdS correlator is, as we have seen in (\ref{eq:GfAdS}), given by 
\begin{equation}
	G_{F}^{\text{AdS}}(v_2,z_2;v_1,z_1;k) =\frac{\sqrt{z_1 z_2}}{2\pi}\left[K_0\left(\sqrt{a_1}\,|k|\right)-K_0\left(\sqrt{a_2}\,|k|\right)\right],
	\label{eq:GfAdS}
\end{equation}
where
\begin{align}
	a_1 &= -(v_2-v_1)^2-2(v_2-v_1)(z_2-z_1)+i\epsilon,\notag\\
	a_2 &= -(v_2-v_1)^2-2(v_2-v_1)(z_2-z_1)+4z_1 z_2+i\epsilon.
\end{align}
The horizon in the AdS-part of Vaidya spacetime is given by the outermost outgoing null geodesics that reach $z=1$ at $v=0$. The outgoing geodesics are parametrized by $v+2z=\text{const}$. Therefore, the horizon is located at $v+2z=2$, i.e.\ for fixed $v_1<0$ and $v_2<0$, the near-horizon limit is
\begin{equation}
	z_1\to 1-\frac{v_1}{2} \qquad\text{and}\qquad z_2\to 1-\frac{v_2}{2},
\end{equation}
where the two limits are approached at the same rate. In the near-horizon limit $a_1\to 0$ while $a_2$ is finite. Therefore, only the first of the Bessel functions contributes to the near-horizon correlator as it diverges logarithmically when $a_1\to 0$ . Therefore, the dominant term is
\begin{equation}
	-\frac{1}{4\pi}\log\left[v_2-v_1+2z_2-2z_1\right].
	\label{eq:aux1}
\end{equation}
This is sufficient to obtain the outgoing part of the initial data. Now, by imposing continuity of the correlator at $v=0$ we obtain\footnote{Strictly speaking the ingoing part $G_{F,ii}(0,0)$ is infinite due to lightlike separation. A more careful version of the calculation would include keeping $v_1$ and $v_2$ separate and requiring continuity of the correlator first for $v_1$ and then separately for $v_2$. This leads to same result (\ref{eq:nearhorizon}) and thus we will follow the quicker route to the result here.}
\beq
	G_{F,oo}(2{z_2}_*,2{z_1}_*)+G_{F,io}(0,2{z_1}_*)+G_{F,oi}(2{z_2}_*,0)+G_{F,ii}(0,0)\approx -\frac{1}{4\pi}\log\left[z_2-z_1\right].
	\label{eq:matchingnearhorizon}
\eeq
The above equation implies that the outgoing part is fixed up to a constant to be
\beq
G_{F,oo}(2{z_2}_*,2{z_1}_*)\approx  -\frac{1}{4\pi}\log\left[z_2-z_1\right].
\eeq
Near the horizon, we can approximate $z\approx 1-2 e^{-2 z_*}$ which gives
\beq
G_{F,oo}(v_2+2{z_2}_*,v_1+2{z_1}_*)\approx  -\frac{1}{4\pi}\log\left[e^{-(v_1+2{z_1}_*)}-e^{-(v_2+2{z_2}_*)}\right].
\label{eq:nearhorizon}
\eeq
A straightforward manipulation of hypergeometric functions shows that the outgoing part of the thermal BTZ propagator (\ref{eq:GfBTZ})
indeed agrees with (\ref{eq:nearhorizon}). Another way of seeing thermality in (\ref{eq:nearhorizon}) is to note that the
correlator is periodic in imaginary time $v\rightarrow v+2\pi i$, recalling that in our units $\beta=2\pi$.

To summarize, we have shown that the outgoing part of the near horizon correlator in the Vaidya spacetime is identical to the outgoing part of the near horizon thermal correlator in the BTZ background. Thus, as the value of the correlator at very late times is only sensitive to the near horizon correlator, we can conclude that the correlator thermalizes at late times. 

\begin{figure}
	\centering
	\includegraphics[width=13.5cm]{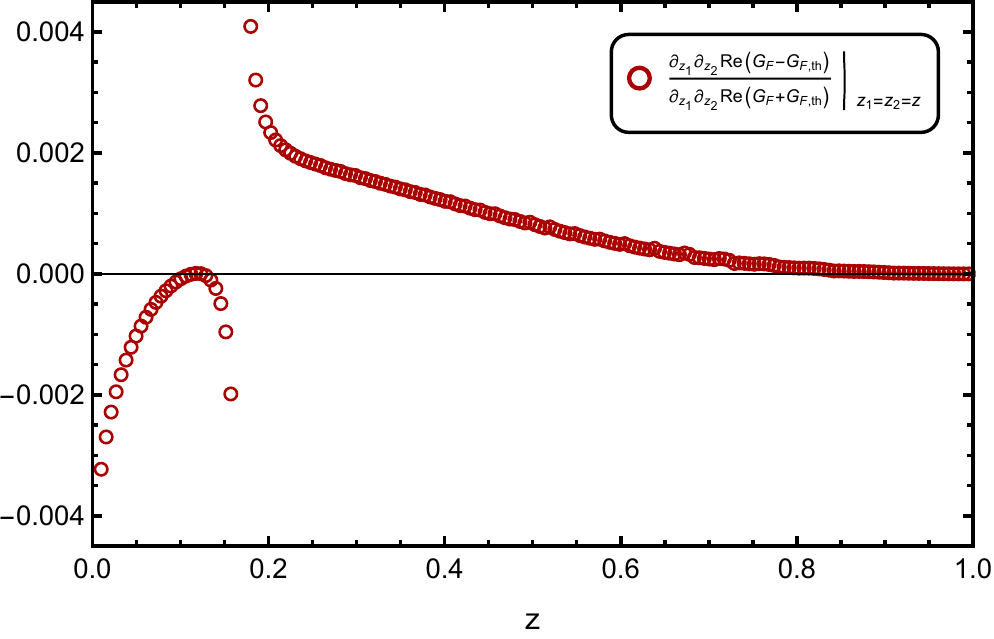}
	\caption{The quantity $\left.\frac{\partial_{z_1}\partial_{z_2}\Re\left[G_F-G_{F,th}\right]}{\partial_{z_1}\partial_{z_2}\Re\left[G_F+G_{F,th}\right]}\right|_{z_1=z_2=z}$ is plotted as a function of $z$ with parameters $v_1=0.01$, $v_2=0.26$ and $k=1$. Finite differences have been employed to approximate the derivatives. We can see that the relative difference between the derivatives of the  Vaidya and the thermal correlator vanishes close to the horizon signalling that the outgoing part of the near-horizon correlator thermalizes quickly behind the shock wave, while it is not thermal closer to the boundary yet.}
	\label{fig:near_horizon_thermalization}
\end{figure}

In Figure \ref{fig:near_horizon_thermalization}, we show that indeed our numerical calculation of the full Vaidya correlator supports the above conclusion. To isolate the outgoing part, we consider the derivative $\partial^2 G_F/\partial z_1\partial z_2$, where the derivatives get rid of the ingoing and mixed terms in the near horizon correlator (\ref{eq:propagator_decomposition}). We compare this quantity to the corresponding thermal one as a function of $z$. Indeed, what we find is that just outside the shock wave the outgoing Vaidya correlator agrees with the thermal correlator with good accuracy near the horizon, while the difference between the two increases further away from the horizon. 

\section{Comparison with geodesic approximation}\label{sec:geodappr}

The bulk to bulk correlator of a free scalar field is given by
an inverse of the operator $\Box-m^2$. 
This inverse operator $(\Box-m^2)^{-1}$ has a well-known path integral representation
as a path integral of a relativistic particle (see e.g.\ \cite{Hartle:1976tp,Louko:2000tp})
\begin{equation}
	(\Box-m^2)^{-1}= \int \left[dx(t)\right]e^{- m\int_0^{s}d\tau |\dot{x}(\tau)|}.\label{eq:1}
\end{equation}
When $m$ is large, we can use a saddle point approximation in the path integral. This way, we obtain 
an approximate expression for the bulk to bulk correlator as
\begin{equation}
	G(x_2;x_1)\propto e^{- m L[x_{cl}]},\label{eq:saddlepoint}
\end{equation}
where $x_{cl}(\tau)$ minimizes the length functional $L=\int_{\tau_1}^{\tau_2}d\tau |\dot{x}_{cl}(\tau)|$, i.e.\ it is a geodesic. This procedure is called the
geodesic approximation.

In Euclidean time, the inverse in (\ref{eq:1}) is unique and the resulting length functional is real and the approximation above can be justified. The case of real time is more subtle, as discussed in \cite{Louko:2000tp}. For real time, the inverse of $\Box-m^2$ is no longer unique as one can add to it any solution of the equation of motion $(\Box-m^2)G=0$. This reflects the fact that in real time, one  must specify the state of the quantum field $\phi$, while in the semiclassical Euclidean setting, the state is fixed by regularity. Even when ignoring this issue, the saddle point approximation of the resulting path integral is in general somewhat problematic. For spacelike geodesics, the action in the exponent is real so a saddle point approximation would seem viable. Recalling that in a Lorentzian spacetime a spacelike geodesic is not a curve of minimal length, it becomes clear that in order to apply a stationary phase approximation, one must rotate the integration contour in the path integral to complex values of the coordinates to reach the stationary phase contour. Thus, one has to assume that the spacetime metric is analytic in order to justify such a rotation. The metric of the Vaidya spacetime is not analytic as a function of $v$. Thus, it is not clear whether the geodesic approximation can be applied in this case. Despite these uncertainties, several works have used the geodesic approximation to calculate correlation functions in the BTZ-Vaidya spacetime, finding physically reasonable results for the boundary correlators \cite{Balasubramanian:2011ur,Aparicio:2011zy}. In this Section, we will present a comparison of how our results for the correlation functions compare to results obtained using the geodesic approximation.

To apply the geodesic approximation, one has to find the spacelike geodesics connecting  the corresponding boundary points. For the BTZ-Vaidya spacetime, this was done in \cite{Balasubramanian:2011ur,Aparicio:2011zy}, where we refer the reader to for details of the calculations. The relevant results we need for the following are summarized in Appendix \ref{sec:geodesicapp}. As we want to compare the thermalization of the geodesic correlators to our results, we need to calculate the Fourier transform
\begin{equation}
	\delta G(t_2;t_1;k)=\int dx\, e^{-ikx}\Big(G(x,t_2;0,t_1)-G_{th}(x,t_2;0,t_1)\Big),
\end{equation}
where $G_{th}(x,t_2;0,t_1)$ is the thermal correlation function, which is the same for the geodesic approximation as for the exact result (\ref{eq:GfBTZ}).\footnote{This follows because the functional form of the boundary correlator is fixed by conformal symmetry, while we choose the coefficient in front of (\ref{eq:saddlepoint}) in a way that the vacuum correlator is canonically normalized to $1/|x|^{2\Delta}$.}

To obtain some intuition about how the quantity $\delta G$ is expected to behave, we will first cook up a toy model for the geodesic correlator. For $|x|<t_1+t_2$ the geodesic correlator is known to be given by the thermal correlator. On the other hand for $|x|\gg t_1+t_2$, it is given by \cite{Aparicio:2011zy}
\begin{equation}
	G(x,t_2;0,t_1)\approx \frac{1}{|x|^{2\Delta}(\cosh(t_1/2)\cosh(t_2/2))^{2\Delta}}.
\end{equation}
Thus, our toy model for the geodesic correlator is
\begin{equation}
	G(x,t_2;0,t_1)= \left\{\begin{array}{cl}
		G_{th}(x,t_2;0,t_1), & |x|<t_1+t_2\, \\
 		\frac{1}{|x|^{2\Delta}(\cosh(t_1/2)\cosh(t_2/2))^{2\Delta}}, & |x|>t_1+t_2 
	\end{array} \right. .
\end{equation}
The Fourier transformed correlator is then simply
\begin{equation}
\delta G(t_2;t_1;k)=\frac{2}{(\cosh(t_1/2)\cosh(t_2/2))^{2\Delta}}\int_{t_1+t_2}^{\infty} dx\,\frac{\cos(kx)}{x^{2\Delta}}.
\end{equation}
Some basic features of the integral can be understood as follows. 
As $t_1$ and $t_2$ increase, the integration region over $x$ shrinks as the lower integration limit is given by $t_1+t_2$.
As the integrand decays as $1/(t_1+t_2)^{2\Delta}$, the whole integral is decaying as a power of $(t_1+t_2)$.
Also since the integrand oscillates, the whole integral can be expected to oscillate as $\cos((t_1+t_2)k)$ up to a phase shift. 
These expectations can be made more precise by evaluating the integral at large $t_1+t_2$,
\begin{equation}
\delta G(t_2;t_1;k)=\frac{2}{(\cosh(t_1/2)\cosh(t_2/2))^{2\Delta}}\Bigg[-\frac{\sin(k(t_1+t_2))}{k(t_1+t_2)^{2\Delta}}+O\Big((t_1+t_2)^{-2\Delta-1}\Big)\Bigg],
\label{eq:toymodel}
\end{equation}
where the result was obtained by integration by parts. This toy model correlator has the property that short distance correlations thermalize before the long distance correlations. In momentum space this statement is a bit less clear than in position space. From (\ref{eq:toymodel}) we see that at very large times the correlations for arbitrary non-vanishing $k$ decay with the same rate in time $e^{-\Delta(t_1+t_2)}(t_1+t_2)^{-2\Delta}$. This is quite different from the position space result. In position space the correlator reaches the thermal value exactly at a time $t_1+t_2=|x|$. This pattern shows up in the rate of oscillation of the Fourier transform, as non-analyticity in real space translates into oscillation in momentum space.

\begin{figure}
	\centering
	\includegraphics[width=13cm]{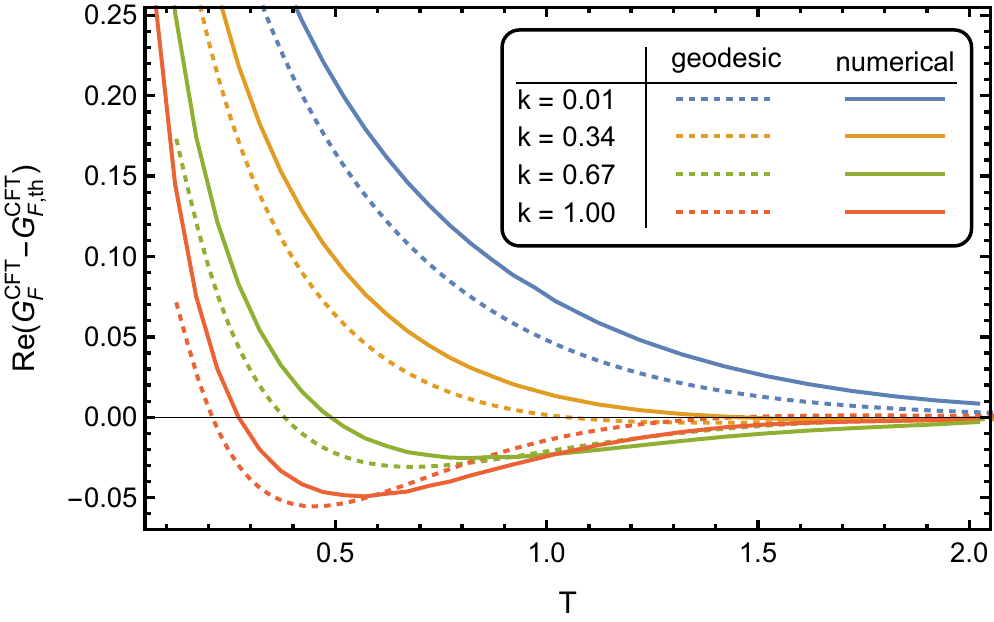}
	\caption{Same as in Figure \ref{fig:thermalization3}. The dashed curves are obtained with our numerical method using (\ref{eq:joining3propak}),
while the solid curves are obtained using the geodesic approximation.}
	\label{fig:geodesic}
\end{figure}

Next, we compare the result of the full geodesic approximation given by first numerically evaluating the correlator in $x$ space using (\ref{eq:geod1}) and then Fourier transforming the corresponding correlator to obtain $\delta G$. Again the integration region starts from $|x|=t_1+t_2$ so that we avoid the problem of having to integrate across the short distance singularity as long as $t_1-t_2\ll t_1+t_2$. We will consider values for $t_1$ and $t_2$ here for which this is true. Results from the geodesic approximation, together with our earlier results, are shown in Figure \ref{fig:geodesic}.  As the figure shows, the geodesic approximation agrees well with our calculation as far as the qualitative features are concerned. Quantitatively, the results generically differ by an overall order one factor.

\section{Conclusions}

In the first part of this work, we studied different versions of the AdS/CFT dictionary for computing out of equilibrium two point correlation functions. In the first version, developed by Skenderis and van Rees in \cite{Skenderis:2008dh,Skenderis:2008dg}, one constructs a holographic version of the Schwinger-Keldysh generating functional. This procedure amounts to calculating the on-shell action for solutions of the bulk equations of motion in a spacetime that is obtained by gluing together Euclidean and Lorentzian spacetimes to construct the Schwinger-Keldysh contour. In this formalism, correlation functions are obtained by taking functional derivatives of the on-shell action. The second version of the AdS/CFT dictionary we discussed was a non-equilibrium version of the "extrapolate" dictionary \cite{Banks:1998dd}. In this dictionary, boundary correlation functions are obtained from the bulk correlation functions with the operator replacement $\mathcal{O}(x)=\lim_{z\rightarrow 0}z^{-\Delta}\phi(x,z)$. In Section \ref{sec:dictproof}, we explicitly showed that the two dictionaries are equivalent, by showing that the bulk to boundary propagators following from the "extrapolate" dictionary satisfy all the equations of motion and boundary conditions following from the SvR prescription. Thus, the two dictionaries give the same "in-in" two point correlation functions in the boundary CFT. We believe that the equivalence might hold beyond two point functions, but we have not yet constructed a proof of this statement. One approach might be to generalize the path integral approach of \cite{Harlow:2011ke} to include real time and external wavefunctions. We will leave this problem for future work.

In the second part of the paper, we studied examples of two point functions in dynamical spacetimes corresponding to sudden quenches of the dual CFT. In this kind of quench, one starts from the ground state of the CFT and suddenly perturbs it out of equilibrium by turning on time dependent sources for some operators. This procedure can be modelled with the Vaidya spacetime. 

In this work, we studied two point correlation functions of scalar fields in the AdS$_2$-Vaidya and AdS$_3$-Vaidya spacetimes. As far as we know, this provides the first study of correlation functions in a collapsing spacetime that starts from the AdS$_d$ vacuum (with $d>2$) and that does not use the geodesic approximation (for correlators in collapse starting from an initial black hole see \cite{Chesler:2011ds,Chesler:2012zk}). We provided a straightforward numerical procedure for calculating the correlation functions in the Vaidya spacetime. To simplify the calculational steps, we considered conformally coupled scalar fields. We do not believe that there is any in principle obstruction for generalizing our calculations to other values of scalar field masses and to other bulk fields. In particular, it would be interesting to study how the correlation functions depend on the mass of the bulk scalar $m$ and see whether at large $m$, the results approach the geodesic approximation. This is a non-trivial question, as the geodesic approximation suffers from several problems discussed in \cite{Louko:2000tp} and in our Section \ref{sec:geodappr}, so that it is not clear whether the approximation is well justified.

For AdS$_2$-Vaidya, the two point correlators for a massless scalar were computed in \cite{Ebrahim:2010ra}. Here, we used this example as a warm-up problem, to show how our calculational procedure works. In this case, one finds that the two point correlator thermalizes immediately as both of the points are evolved past the collapsing shell. 

For AdS$_3$-Vaidya, we found that the Feynman two point function is approaching the thermal two point function. This approach was found to depend on the momentum $k$. In particular, we found that for small $k$, the two point function approaches the thermal value slower than for larger $k$. This is consistent with the general pattern found from many different systems, that  longer distance correlations thermalize slower. This has been observed in the geodesic approximation, in the AdS$_3$-Vaidya spacetime \cite{Balasubramanian:2011ur,Aparicio:2011zy} as well as in CFT calculations in certain type of quenches \cite{Calabrese:2006rx} and even in cold atom experiments \cite{Langen2013}.

In more detail, we compared our results for the two point function to the Fourier transformed geodesic two point functions. The basic qualitative features were found to agree, as can be seen in Figure \ref{fig:geodesic}.

\section*{Acknowledgements}

We would like to thank Jorge Casalderrey-Solana, Ben Craps, Dimitrios Giataganas, Esko Keski-Vakkuri, Andy O'Bannon, 
Mukund Rangamani, Andrei Starinets, Stefan Stricker, Olli Taanila and Aleksi Vuorinen for useful discussions. 
This research was supported by the European Research Council under the European Union's Seventh Framework 
Programme (ERC Grant agreement 307955). VK would like to thank the Helsinki Institute of Physics for hospitality while 
this work was in progress.

\newpage
\section*{Appendix}
\appendix

\section{Identities between two point functions}
\label{sec:correlatoridentities}

The Feynman correlator is given by
\beq
G_F(x_2,x_1)=\langle T\phi(x_2)\phi(x_1)\rangle=\theta(t_1-t_2)G_+(x_1,x_2)+\theta(t_2-t_1)G_+(x_1,x_2)^*,\label{eq:feynman}
\eeq
where we defined the Wightman function
\beq
G_+(x_1,x_2)=\langle \phi(x_1)\phi(x_2)\rangle,
\eeq
and we used
\beq
G_+(x_1,x_2)^*=\langle \phi(x_1)\phi(x_2)\rangle^*=\langle \phi(x_2)\phi(x_1)\rangle=G_+(x_2,x_1).
\eeq
The retarded correlator is given by
\beq
G_R(x_2,x_1)=\theta(t_2-t_1)\langle [\phi(x_2),\phi(x_1)]\rangle=\theta(t_2-t_1)(G_+(x_2,x_1)-G_+(x_2,x_1)^*),
\eeq
which can be compactly written as
\beq
G_R(x_2,x_1)=2i\theta(t_2-t_1)\textrm{Im} G_+(x_2,x_1).
\eeq
Using (\ref{eq:feynman}), we also get the relation
\beq
G_R(x_2,x_1)=2i\theta(t_2-t_1)\textrm{Im} G_F(x_2,x_1).\label{eq:retarded}
\eeq
Another convenient relation is 
\beq
G_F(x_2,x_1)=G_+(x_1,x_2)+G_R(x_2,x_1).
\eeq
On the other hand the Wightman function $G_+$ can be obtained from the
Feynman correlator as
\beq
G_+(x_2,x_1)=\theta(t_2-t_1)G_F(x_2,x_1)+\theta(t_1-t_2)G_F^*(x_2,x_1).
\eeq

\section{Initial state path integrals}\label{sec:pathintegrals}

We need the following expectation values
\beq
\left.\langle \Pi(x_1)\phi(x_2)\rangle\right|_{t_1=t_2} = \int [d\phi_i]\Psi^*[\phi_i]\Big(-i\frac{\delta}{\delta\phi_i(\bm{x}_1)}\phi_i(\bm{x}_2)\Big)\Psi[\phi_i],
\eeq
where $\Psi$ is the wavefunctional given in (\ref{eq:wavefunctional}). Taking the functional derivatives gives
\begin{align}
\langle \left.\Pi(x_1)\phi(x_2)\rangle\right|_{t_1=t_2} &= -i\delta(\bm{x}_1-\bm{x}_2)+\frac{i}{2\mathcal{N}}\int d\bm{x}\int [d\phi_i]e^{-\frac{1}{2}\int \phi_i K\phi_i}K(\bm{x},\bm{x}_1)\phi_i(\bm{x})\phi_i(\bm{x}_2)\notag
\\
&= -i\delta(\bm{x}_1-\bm{x}_2)+\frac{i}{2}\int d\bm{x}\, K(\bm{x},\bm{x}_1)G(\bm{x},\bm{x}_2)\notag
\\
&=-\frac{i}{2}\delta(\bm{x}_1-\bm{x}_2).
\end{align}
On the other hand
\beq
\langle \left.\phi(x_1) \Pi(x_2)\rangle\right|_{t_1=t_2}=\frac{i}{2}\int d\bm{x}\, K(\bm{x},\bm{x}_2)G(\bm{x},\bm{x}_1)=\frac{i}{2}\delta(\bm{x}_1-\bm{x}_2).
\eeq
Using these two results we also get
\begin{align}
&\left. D^{t_1} \langle T \phi(x_1)\phi(x_2)\rangle\right|_{t_2=t_1+\epsilon} \notag
\\
&=-\theta(t_1-t_2)\langle \left.\Pi(x_1)\phi(x_2)\rangle\right|_{t_2=t_1}-\theta(t_2-t_1)\left.\langle \phi(x_2)\Pi(x_1)\rangle\right|_{t_2=t_1}
+O(\epsilon)\notag
\\
&=\textrm{sign}(t_1-t_2) \frac{i}{2}\delta(\bm{x}_1-\bm{x}_2)+O(\epsilon),\label{eq:epsilon1}
\end{align}
where we used $D^t\phi=-\Pi$. Another useful expectation value is
\begin{align}
&\left.\langle \Pi(x_1)\Pi(x_2)\rangle\right|_{t_1=t_2}=
\frac{1}{\mathcal{N}}\int [d\phi_i]\Psi^*[\phi_i]\Big(-i\frac{\delta}{\delta\phi_i(\bm{x}_1)}\Big)\Big(-i\frac{\delta}{\delta\phi_i(\bm{x}_2)}\Big)\Psi[\phi_i]
\notag
\\
&=\frac{1}{2}K(\bm{x}_1,\bm{x}_2)-\frac{1}{4\mathcal{N}}\int d\bm{x} d\bm{y}\,\int [d\phi_i]e^{-\frac{1}{2}\int \phi_i K\phi_i}K(\bm{x},\bm{x}_1)\phi_i(\bm{x})K(\bm{y},\bm{x}_2)\phi_i(\bm{y})
\notag
\\
&=\frac{1}{2}K(\bm{x}_1,\bm{x}_2)-\frac{1}{4}\int d\bm{x} d\bm{y}\,K(\bm{x},\bm{x}_1)K(\bm{y},\bm{x}_2)G(\bm{x},\bm{y})\notag
\\
&=\frac{1}{4}K(\bm{x}_1,\bm{x}_2).
\end{align}
Using this result, we obtain
\begin{align}
\left.D^{t_1}D^{t_2}\langle \phi(x_1)\phi(x_2)\rangle\right|_{t_2=t_1+\epsilon} &= \frac{1}{4}K(\bm{x}_1,\bm{x}_2)+O(\epsilon),
\label{eq:epsilon2}
\\
\left. D^{t_1}D^{t_2}\langle T \phi(x_1)\phi(x_2)\rangle\right|_{t_2=t_1+\epsilon}&=\frac{1}{4}K(\bm{x}_1,\bm{x}_2)-iD^{t_2}\theta(t_2-t_1)\delta(\bm{x}_1-\bm{x}_2)+O(\epsilon).\label{eq:epsilon3}
\end{align}
The order $\epsilon$ terms in (\ref{eq:epsilon1}), (\ref{eq:epsilon2}) and (\ref{eq:epsilon3}), can be systematically computed as follows.
For a general operator $A$, one can write
\beq
A(t_2)=U^{{\dagger}}(t_2,t_1)A(t_1)U(t_2,t_1)=A(t_1)+i\epsilon [H,A(t_1)]+O(\epsilon^2).
\eeq
Then, the commutators can be determined from the Heisenberg equations of motion. We will not perform this calculation
here as we do not need the $O(\epsilon)$ terms explicitly.

\section{Calculational details on the AdS$_3$-Vaidya correlator}\label{sec:details}

Following an approach from \cite{Balasubramanian:2012tu} we compute the Feynman propagator on the BTZ-side of the Vaidya spacetime by time evolution of the propagator on the AdS-side. This is done in two steps:
\begin{enumerate}
	\item{Join two propagators on the AdS- and BTZ-side of the Vaidya spacetime to get the Feynman propagator across the shell.}
	\item{Evolve the Feynman propagator across the shell to get the propagator on the BTZ side of the spacetime.}
\end{enumerate}
These two steps can be understood as performing the integrals in (\ref{eq:joining3propak}) in two steps. The integrals over $z_1$ and $z_2$ yield propagators across the shell. The final integration over $z_0$ gives the Feynman propagator fully on the BTZ side of the geometry.

\subsection{Step 1: Feynman propagator across the shell}

As shown in \cite{Balasubramanian:2012tu}, the Feynman propagator across the shell in Vaidya spacetime, i.e.\ for $v_0<0$ and $v_1>0$, can be computed as
\begin{align}
	G_{F}(v_1,z_1;v_0,z_0;k)=&\,i\int_{v=0}dz\,\sqrt{-g}g^{v\mu}\left(G_{F}^\text{AdS}(0,z;v_0,z_0;k)\overleftrightarrow{\partial_\mu}G_{R}^\text{BTZ}(v_1,z_1;0,z;k)\right)\notag\\
		=&-i\int_{v=0}\frac{dz}{z}\left(G_{F}^\text{AdS}(0,z;v_0,z_0;k)\overleftrightarrow{\partial_z}G_{R}^\text{BTZ}(v_1,z_1;0,z;k)\right)\notag\\
		=& -i\int_{v=0}\frac{dz}{z}\bigg(2G_{F}^\text{AdS}(0,z;v_0,z_0;k)\partial_z G_{R}^\text{BTZ}(v_1,z_1;0,z;k)\notag\\
		&	\qquad\qquad\quad -\frac{1}{z}G_{F}^\text{AdS}(0,z;v_0,z_0;k) G_{R}^\text{BTZ}(v_1,z_1;0,z;k)\bigg)\notag\\
		&\left. +\frac{i}{z}G_{F}^\text{AdS}(0,z;v_0,z_0;k) G_{R}^\text{BTZ}(v_1,z_1;0,z;k)\right|_{z=0}^{z=1}\notag\\
		=& -i\int_{v=0}\frac{dz}{\sqrt{z}}\,2G_{F}^\text{AdS}(0,z;v_0,z_0;k)\partial_z \left(\frac{1}{\sqrt{z}}G_{R}^\text{BTZ}(v_1,z_1;0,z;k)\right)
	\label{eq:Gf_unextracted}
\end{align}
where we have used integration by parts with the boundary term vanishing. We can restrict the integration region to the interval $\left[0,1\right]$ as the retarded correlator $G_{R}^\text{BTZ}(v_1,z_1;0,z;k)$ vanishes for $z>1$, i.e.\ for $(v=0,z)$ behind the horizon, due to causality. As our initial state is the pure AdS vacuum, we know the Feynman correlator in \eqref{eq:Gf_unextracted}. The same is true for the retarded correlator on the BTZ-side, since the retarded correlator for a free field does not depend on the quantum state after the collapse (whereas the Feynman correlator does such that we cannot infer it from pure BTZ). Since we have computed the pure AdS and BTZ correlators in \eqref{eq:GfAdS} and \eqref{eq:GfBTZ}, we can compute the integral numerically. However, we need to extract $\delta$-function contributions coming from the derivative of the retarded correlator on the BTZ side. The retarded correlator is
\begin{equation}
	G_{R}^\text{BTZ}(v_1,z_1;0,z;k)=2i\Im G_{F}^\text{BTZ}(v_1,z_1;0,z;k).
\end{equation}
$G_{F}^\text{BTZ}$ has logarithmic divergences at $b_1(v_1,z_1;0,z)=-1$ and $b_2(v_1,z_1;0,z)=-1$, where $(v_1,z_1)$ and $(v_0,z_0)$ are lightlike separated (see (\ref{eq:b1_and_b2}) for the definition of $b_1$ and $b_2$). At these points, the imaginary part behaves like a Heaviside step function. Consequently, the derivative contributes a $\delta$-function to the integrand. Since this $\delta$-function is not taken into account by numerical integration, we compute it analytically:
\begin{align}
	\partial_z 2i\Im G_{F}^\text{BTZ}&(v_1,z_1;0,z;k)\notag\\
		&\xrightarrow[\text{part}]{\text{div.}} i\partial_z \Im \frac{1}{2\pi}\sqrt{\frac{z z_1}{4\pi}}\left[-2\sqrt{\pi}\log(1+b_1(v_1,z_1;0,z))+2\sqrt{\pi}\log(1+b_2(v_1,z_1;0,z))\right]\notag\\
		&\longrightarrow\frac{i\sqrt{z z_1}}{2\pi}\partial_z \left[-\pi \theta(-1-b_1(v_1,z_1;0,z))+\pi \theta(-1-b_2(v_1,z_1;0,z))\right]\notag\\
		&= \frac{i\sqrt{z z_1}}{2}\left[\d\left(z-z_+(v_1,z_1)\right)-\d\left(z-z_-(v_1,z_1)\right)\right],
\end{align}
where $z_+(v_1,z_1)$ and $z_-(v_1,z_1)$ are the solutions of $b_1(v_1,z_1;0,z)=-1$ and $b_2(v_1,z_1;0,z)=-1$ respectively,
\begin{align}
	z_+(v_1,z_1)&=\frac{-1+\cosh v_1+z_1 \sinh v_1}{-z_1+z_1 \cosh v_1 +\sinh v_1},\notag\\
	z_-(v_1,z_1)&=\tanh\frac{v_1}{2}.
	\label{eq:z_plus_minus}
\end{align}
$z_+(v_1,z_1)$ is the upper boundary for the $z$-integral in \eqref{eq:Gf_unextracted} since the retarded correlator $G_{R}^\text{BTZ}(v_1,z_1;0,z;k)$ vanishes for $z$ outside of the backward lightcone. Therefore, we can conclude that
\begin{align}
	G_{F}(v_1,z_1;v_0,z_0;k)=&-i\int'\frac{dz}{\sqrt{z}}\,2G_{F}^\text{AdS}(0,z;v_0,z_0;k)\partial_z \left(\frac{1}{\sqrt{z}}G_{R}^\text{BTZ}(v_1,z_1;0,z;k)\right)\notag\\
		&+\sqrt{\frac{z_1}{z_+(v_1,z_1)}}G_{F}^\text{AdS}(0,z_+(v_1,z_1);v_0,z_0;k)\notag\\
		&-\sqrt{\frac{z_1}{z_-(v_1,z_1)}}G_{F}^\text{AdS}(0,z_-(v_1,z_1);v_0,z_0;k)\notag\\
		\equiv&\, G_{F,\text{num}}^a(v_1,z_1;v_0,z_0;k)+G_{F,\text{ana}}^a(v_1,z_1;v_0,z_0;k),
	\label{eq:Gf}
\end{align}
where the dash indicates that the $z$-integral has the range from zero to $z_+(v_1,z_1)$ and is not including $z_-(v_1,z_1)$ and $z_+(v_1,z_1)$. $G_{F,\text{num}}^a$ and $G_{F,\text{ana}}^a$ are the parts of the propagator that are known numerically and analytically respectively,
\begin{align}
	G_{F,\text{num}}^a(v_1,z_1;v_0,z_0;k)\equiv&-i\int'\frac{dz}{\sqrt{z}}\,2G_{F}^\text{AdS}(0,z;v_0,z_0;k)\partial_z \left(\frac{1}{\sqrt{z}}G_{R}^\text{BTZ}(v_1,z_1;0,z;k)\right),
	\label{eq:Gfanum}\\
	G_{F,\text{ana}}^a(v_1,z_1;v_0,z_0;k)\equiv&\,\sqrt{\frac{z_1}{z_+(v_1,z_1)}}G_{F}^\text{AdS}(0,z_+(v_1,z_1);v_0,z_0;k)\notag\\
		&-\sqrt{\frac{z_1}{z_-(v_1,z_1)}}G_{F}^\text{AdS}(0,z_-(v_1,z_1);v_0,z_0;k),
	\label{eq:Gfaana}
\end{align}
where the superscript $a$ indicates that these are parts of the correlator across the shell.

From the Feynman correlator we can extract the retarded correlator and its derivative which we will require in the next step,
\begin{equation}
	G_{R}^a(v_1,z_1;v_0,z_0;k) = 2i\Im G_{F}^a(v_1,z_1;v_0,z_0;k)
	\label{eq:Gr}
\end{equation}
and
\begin{align}
	\partial_{z_0} G_{R}^a(v_1,z_1;v_0,z_0;k) = &\: 2i\Im\bigg\{-i\int'\frac{dz}{\sqrt{z}}\left[2\partial_{z_0}G_{F}^\text{AdS}(0,z;v_0,z_0;k)\partial_z \left(\frac{1}{\sqrt{z}}G_{R}^\text{BTZ}(v_1,z_1;0,z;k)\right)\right]\notag\\
		&\qquad\quad+\sqrt{\frac{z_1}{z_+(v_1,z_1)}}\partial_{z_0}G_{{F}}^\text{AdS}(0,z_+(v_1,z_1);v_0,z_0;k)\notag\\
		&\qquad\quad-\sqrt{\frac{z_1}{z_-(v_1,z_1)}}\partial_{z_0}G_{{F}}^\text{AdS}(0,z_-(v_1,z_1);v_0,z_0;k)\bigg\}\notag\\
		&+\left.\sqrt{z_0}\,\theta\left(z_0+\frac{v_0}{2}\right)\partial_z\left(\frac{1}{\sqrt{z}}G_{R}^\text{BTZ}(v_1,z_1;0,z;k)\right)\right|_{z=z_0+\frac{v_0}{2}}\notag\\
		&+\frac{i\sqrt{z_0 z_1}}{2}\left[\d\left(z_0-\left(z_+(v_1,z_1)-\frac{v_0}{2}\right)\right)-\d\left(z_0-\left(z_-(v_1,z_1)-\frac{v_0}{2}\right)\right)\right]\notag\\
		\equiv \:& \left(\partial_{z_0}G_{R}\right)_\text{reg}(v_1,z_1;v_0,z_0;k)\notag\\
		&+\frac{i\sqrt{z_0 z_1}}{2}\left[\d\left(z_0-\left(z_+(v_1,z_1)-\frac{v_0}{2}\right)\right)-\d\left(z_0-\left(z_-(v_1,z_1)-\frac{v_0}{2}\right)\right)\right].
	\label{eq:DGr}
\end{align}
To get this result we had to extract a $\delta$-function contribution from the derivative of the imaginary part of the AdS-propagator inside the $z$-integral which gives rise to the fourth term which only contributes for $z_0+v_0/2>0$. The AdS-propagator has another delta function contribution $\delta(-z_0-v_0/2)$, which is independent of $z$, and can be taken out of the integral. The remaining integral is a total derivative, and can be seen to vanish. The last two terms come from the logarithmic divergences of the AdS-propagators outside the $z$-integral. We write $\left(\partial_{z_0}G_{R}\right)_\text{reg}$ for $\partial_{z_0} G_{R}$ with all $\delta$-function divergences extracted.

\subsection{Step 2: Evolve the Feynman propagator to the BTZ side}

We can finally compute the Feynman propagator on the BTZ side of the Vaidya spacetime given by
\begin{equation}
	G_{F}(v_2,z_2;v_1,z_1;k) = -i\int_{v=v_0}\frac{dz_0}{\sqrt{z_0}}\left[2G_{F}^a(v_2,z_2;v_0,z_0;k) \partial_{z_0}\left(\frac{1}{\sqrt{z_0}}G_{R}^a(v_1,z_1;v_0,z_0;k)\right)\right]
\end{equation}
for some $v_1,v_2>0$ and $v_0<0$. Using the results from \eqref{eq:Gf}, \eqref{eq:Gr} and \eqref{eq:DGr}, we can express the above integral with all $\d$-functions extracted,
\begin{align}
	G_{F}(v_2,z_2;v_1,z_1;k) =& -i\int'\frac{dz_0}{z_0}\bigg[2G_{F}^a(v_2,z_2;v_0,z_0;k)\bigg(\left(\partial_{z_0}G_{R}^a\right)_\text{reg}(v_1,z_1;v_0,z_0;k)\notag\\
			& \qquad\qquad\qquad -\frac{1}{2z_0} G_{R}^a(v_1,z_1;v_0,z_0;k)\bigg)\bigg]\notag\\
			& +\sqrt{\frac{z_1}{z_+(v_1,z_1)-\frac{v_0}{2}}}\,2G_{F}^a\left(v_2,z_2;v_0,z_+(v_1,z_1)-\frac{v_0}{2};k\right)\notag\\
			& -\sqrt{\frac{z_1}{z_-(v_1,z_1)-\frac{v_0}{2}}}\,2G_{F}^a\left(v_2,z_2;v_0,z_-(v_1,z_1)-\frac{v_0}{2};k\right).
\end{align}
The $z_0$-integral excludes the points $\{z_+(v_1,z_1)-v_0/2,z_-(v_1,z_1)-v_0/2\}$ at which we extracted the $\d$-function contributions. Using the expression \eqref{eq:Gf} for the propagator across the shell we can split the Feynman propagator into numerical and analytical part,
\begin{equation}
	G_{F}(v_2,z_2;v_1,z_1;k) = G_{F,\text{num}}(v_2,z_2;v_1,z_1;k)+G_{F,\text{ana}}(v_2,z_2;v_1,z_1;k),
	\label{eq:G}
\end{equation}
using the expressions for the numerical and analytical parts of the propagator across the shell from (\ref{eq:Gfanum}) and (\ref{eq:Gfaana}). We obtain
\begin{align}
	G_{F,\text{num}}(v_2,z_2;v_1,z_1;k) = & -i\int'\frac{dz_0}{z_0}\bigg[2G_{F}^a(v_2,z_2;v_0,z_0;k)\bigg(\left(\partial_{z_0}G_{R}^a\right)_\text{reg}(v_1,z_1;v_0,z_0;k)\notag\\
			& \qquad\qquad\qquad -\frac{1}{2z_0} G_{R}^a(v_1,z_1;v_0,z_0;k)\bigg)\bigg]\notag\\
			&+\sqrt{\frac{z_1}{z_+(v_1,z_1)-\frac{v_0}{2}}}G_{F,\text{num}}^{a}\left(v_2,z_2;v_0,z_+(v_1,z_1)-\frac{v_0}{2};k\right)\notag\\
			&-\sqrt{\frac{z_1}{z_-(v_1,z_1)-\frac{v_0}{2}}}G_{F,\text{num}}^{a}\left(v_2,z_2;v_0,z_-(v_1,z_1)-\frac{v_0}{2};k\right),
	\label{eq:Gnum}
\end{align}
and
\begin{align}
	G_{F,\text{ana}}&(v_2,z_2;v_1,z_1;k)\notag\\
		=& \sqrt{\frac{z_1}{z_+(v_1,z_1)-\frac{v_0}{2}}}G_{F,\text{ana}}^{a}\left(v_2,z_2;v_0,z_+(v_1,z_1)-\frac{v_0}{2};k\right)\notag\\
			& -\sqrt{\frac{z_1}{z_-(v_1,z_1)-\frac{v_0}{2}}}G_{F,\text{ana}}^{a}\left(v_2,z_2;v_0,z_-(v_1,z_1)-\frac{v_0}{2};k\right)\notag\\
		=& \sqrt{\frac{z_1}{z_+(v_1,z_1)-\frac{v_0}{2}}}\bigg[\sqrt{\frac{z_2}{z_+(v_2,z_2)}}G_{F}^\text{AdS}\left(0,z_+(v_2,z_2);v_0,z_+(v_1,z_1)-\frac{v_0}{2};k\right)\notag\\
			& \qquad\qquad\qquad\qquad\quad-\sqrt{\frac{z_2}{z_-(v_2,z_2)}}G_{F}^\text{AdS}\left(0,z_-(v_2,z_2);v_0,z_+(v_1,z_1)-\frac{v_0}{2};k\right)\bigg]\notag\\
		& -\sqrt{\frac{z_1}{z_-(v_1,z_1)-\frac{v_0}{2}}}\bigg[\sqrt{\frac{z_2}{z_+(v_2,z_2)}}G_{F}^\text{AdS}\left(0,z_+(v_2,z_2);v_0,z_-(v_1,z_1)-\frac{v_0}{2};k\right)\notag\\
			& \qquad\qquad\qquad\qquad\quad-\sqrt{\frac{z_2}{z_-(v_2,z_2)}}G_{F}^\text{AdS}\left(0,z_-(v_2,z_2);v_0,z_-(v_1,z_1)-\frac{v_0}{2};k\right)\bigg].
	\label{eq:Gana}	
\end{align}

While the evaluation of (\ref{eq:Gana}) is straightforward as all the functions are known analytically, we want to comment on the numerical method we use to compute (\ref{eq:Gnum}). The second and third terms are single integrals specified in (\ref{eq:Gfanum}). We evaluate them in Mathematica using the NIntegrate command and Exclusions for the points where the logarithmic divergences occur. The first term is more involved since we have to integrate the functions $G_F^a$ and $G_R^a$ which are themselves not known analytically. We evaluate the integrals contributing to $G_F^a$ and $G_R^a$ numerically for $N$ different values of $z_0$ with equal spacing and upper bound $z_+(v_1,z_1)-v_0/2$ as higher values do not contribute to the integral due to causality. Next, we use the command Interpolation to fit a function to the $N$ points. Then we can use the resulting interpolating functions to integrate over $z_0$. This integral is again performed with the command NIntegrate including Exclusions to treat the logarithmic divergences. The numerical error and its dependence on $N$ is studied in Appendix \ref{sec:numerics}.

\section{Estimating the numerical error}\label{sec:numerics}

\begin{figure}
\centering
\includegraphics[scale=.6]{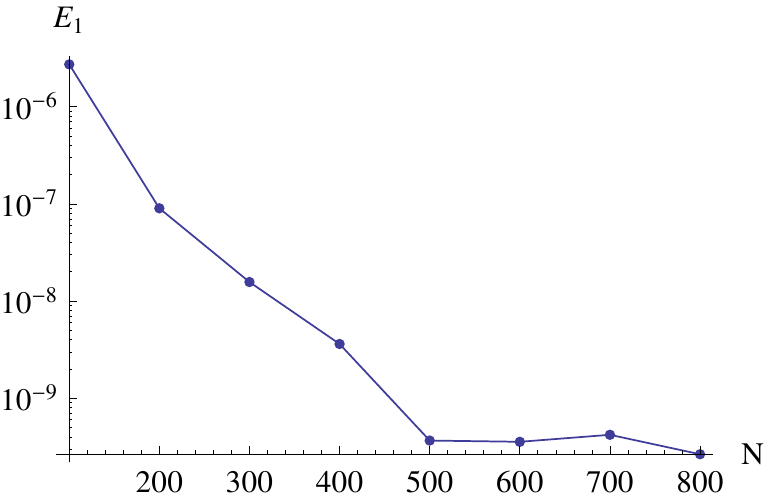}
\includegraphics[scale=.6]{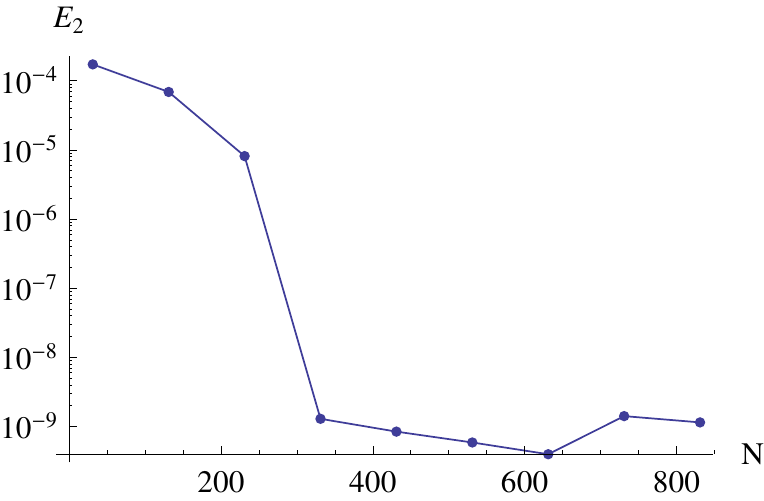}
\includegraphics[scale=.6]{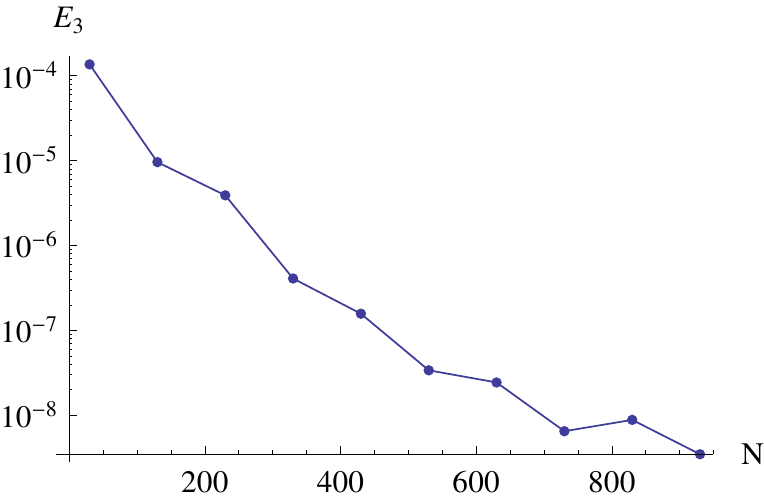}
\caption{\label{error} Different measures of numerical error shown as functions of the number of discretization steps $N$,
corresponding to the parameter values $k=0.5$, $z_2=0.02$, $z_1=0.01$, $v_2=0.5$ and $v_1=0.49$ for $E_1$ and  $k=0.75$, $z_1=0.05$, $z_2=0.055$, $v_2=0.755$ and $v_1=0.75$ for $E_2$ and $E_3$.}
\end{figure}

In this Appendix, we estimate how accurate our numerical procedure for solving the AdS$_3$-Vaidya propagator is. 
As a first check, we can apply the calculational method to evolve the AdS$_3$ propagator forwards in time. 
In this case, we know the exact result analytically, which allows us to test the numerical accuracy of our method.
 In our numerical procedure, we have two sources of error. The first one appears from the  numerical
accuracy of Mathematica routine NIntegrate where we use Mathematica's Machineprecision, which is of the order $10^{-15}$. 
Another source of error appears as we discretize the $z$ space to perform the final integral over $z_0$ in (\ref{eq:joining3propak}). 
The number of points we use in the discretization is denoted by $N$.
The first measure of error we use is
\beq
E_1=\frac{|G_{F}^{\text{AdS},\,\textrm{numerical}}-G_F^{\text{AdS},\,\textrm{exact}}|}{|G_F^{\text{AdS},\,\textrm{exact}}|}.
\eeq
The above measure is plotted in Figure \ref{error} as a function of the number of steps $N$ for a typical values of the parameters specified in the caption. 
The error is seen to decay approximately exponentially until it reaches a plateau at order
$10^{-10}$. From Figure \ref{error}, we see that $N$ controls
the numerical accuracy until at some point we reach a plateau, where increasing $N$ does not help anymore. We believe
that this plateau appears when the error from NIntegrate becomes larger than the discretization error. When
defining the error measure $E_1$, we are dividing by the value of the exact correlation function which is by itself of the
order $10^{-2}$, thus amplifying the error from Machineprecision. 

As the two point function in AdS$_3$-Vaidya is not known analytically, we do not have an obvious
quantity to compare it to. The imaginary part of $G_F$ is expected to thermalize immediately after
both points of the correlation function are in the region $v>0$. Thus, as a second estimate
of the numerical error we use
\beq
E_2=\Bigg|\frac{\textrm{Im}(G_{F}^{\textrm{numerical}}-G_{F,th})}{\textrm{Im}(G_{F,th})}\Bigg|,
\eeq
where now $G_F^{\textrm{numerical}}$ is the correlation function in the Vaidya spacetime. This error estimate
is shown in Figure \ref{error} as a function of the number of steps $N$. We see the same effect here as in the estimate $E_1$.
The error first decays until it reaches a plateau at order $10^{-9}$. Again, we believe that the plateau appears for 
the same reason, namely due to the error in Mathematica's NIntegrate. One should note that here the value of $\textrm{Im}(G_{F,th})$
is of the order $10^{-6}$ which explains why the plateau in $E_2$ is larger than that of $E_1$ by a few orders of magnitude.

As a third measure of the numerical error, we consider
\beq
E_3=\Bigg|\frac{G_{F}^{\textrm{numerical}}(N)}{G_{F}^{\textrm{numerical}}(N+1)}-1\Bigg|,
\eeq
where we displayed the dependence of $G_F$ on the number of steps explicitly. This estimate tells us
how fast the discretized result is converging to a continuum result. We see here that there is no 
clear plateau appearing in this measure. Thus, the discretized integral is approaching a continuum limit,
which as $E_2$ implies, is not quite the right continuum result, but differs from it by an order
of magnitude given by the error in NIntegrate. 

\section{Summary of geodesic approximation results}
\label{sec:geodesicapp}

Let us begin from the results for the equal time correlators in \cite{Balasubramanian:2011ur}. The regulated geodesic length is given by
\begin{equation}
L=2\log\Big(\frac{\sinh(t)}{s(|x|,t)}\Big),
\end{equation}
where the function $s$ is given implicitly by
\begin{equation}
|x|=\frac{2c}{s\rho}+\log\Big(\frac{2(1+c)\rho^2+2s\rho-c}{2(1+c)\rho^2-2s\rho-c}\Big)
\end{equation}
and 
\begin{equation}
\rho=\frac{1}{2}\coth(t)+\frac{1}{2}\sqrt{\coth^2(t)-\frac{2c}{c+1}}
\end{equation}
with $c=\sqrt{1-s^2}$. In our numerical approach, we are forced to work with unequal time correlators and thus, we should see how the above result changes when the boundary times $t_1$ and $t_2$ are slightly different. The geodesics for this case were calculated in full generality (at spacelike separation) in \cite{Aparicio:2011zy}. Here, we will perform a simpler calculation by assuming that $\delta t$ is small, which is sufficient for our purposes. We will denote $t_1=t$ and $t_2=t+\delta t$. Changing the endpoint of the geodesic by a small amount $\delta t$ will induce a change in the geodesic
\begin{equation}
(x_0(z),v_0(z))\rightarrow (x_0(z)+\delta x(z), v_0(z)+\delta v(z)),
\end{equation}
where $(x_0(z),v_0(z))$ denotes the equal time geodesic parametrized as a function of $z$. To first order, the length of the geodesic changes by a boundary term
\begin{equation}
\delta L=\frac{\partial \mathcal{L}}{\partial v'}\delta t,
\end{equation}
where $\mathcal{L}$ is defined through $L=\int dz\,\mathcal{L}$. Using the results in \cite{Balasubramanian:2011ur}, this can be written as\footnote{In more detail, the change in the length in the notation of \cite{Balasubramanian:2011ur}, is $\delta L=-(r^2-1) \dot{t}\,\delta t|_{\lambda\rightarrow\infty}=-\delta t\, E$, where in the second step we used equation (30) of \cite{Balasubramanian:2011ur}. Furthermore by using equations (97) and (100) of \cite{Balasubramanian:2011ur}, we obtain (\ref{eq:deltatee}).}
\begin{equation}
\delta L=\frac{c}{2\rho}\delta t.\label{eq:deltatee}
\end{equation}
Thus, for unequal times but for small time separation, we obtain
\begin{align}
L&=2\log\Big(\frac{\sinh(t_1)}{s}\Big)+\frac{c}{2\rho}(t_2-t_1),\nonumber
\\
|x|&=\frac{2c}{s\rho}+\log\Big(\frac{2(1+c)\rho^2+2s\rho-c}{2(1+c)\rho^2-2s\rho-c}\Big),\label{eq:geod1}
\\
\rho&=\frac{1}{2}\coth(t_1)+\frac{1}{2}\sqrt{\coth^2(t_1)-\frac{2c}{c+1}}.\nonumber
\end{align}

\bibliographystyle{utphys}
\bibliography{references}

\end{document}